\documentclass[11pt]{article}
\usepackage{fullpage,amsmath,amssymb,mathtools,natbib}
\usepackage{graphicx,bbm}
\usepackage{enumerate}
\usepackage[plain]{algorithm2e}
\usepackage{tikz}
\usetikzlibrary{arrows, decorations.markings}

\def\pconv{\smash{\mathop{\longrightarrow}\limits^p}}     
\def\asconv{\smash{\mathop{\longrightarrow}\limits^{a.s.}}} 
\def\dconv{\smash{\mathop{\longrightarrow}\limits^d}}     

\def\cov{\mbox{cov}}
\def\var{\mbox{var}}

\def\argmax{\mbox{argmax}}
\def\argmin{\mbox{argmin}}
\renewcommand{\bar}{\overline}

\renewcommand{\tilde}{\widetilde}
\renewcommand{\hat}{\widehat}

\newcommand{\beq}{\begin{eqnarray*}}
\newcommand{\eeq}{\end{eqnarray*}}

\def\1T{frac{1}{T}}
\def\1n{\frac{1}{n}}

\def\bitem{\medskip\begin{itemize} \itemsep=8.0pt \parskip=8.0pt}
\def\eitem{\end{itemize}}

\newtheorem{proposition}{Proposition}
\newtheorem{lemma}{Lemma}
\def\mymathcal{\mathbbm}

\title{The ABC of Simulation Estimation with Auxiliary Statistics}
\author{Jean-Jacques Forneron\thanks{Department of Economics, Columbia
University. Email: jmf2209@columbia.edu} \and
Serena Ng\thanks{Department of Economics, Columbia
University, and NBER. Email Serena.Ng at Columbia.edu.
\newline Correspondence Address: 420 W. 118 St. Room 1117, New York, NY 10025.
\newline   Financial support is provided by the
 National Science Foundation (SES-0962431 and SES-1558623).
 We thank Richard Davis for discussions that initiated this research, Neil
Shephard, Christopher
 Drovandi, two anonymous referees, and the editors for many helpful suggestions. Comments from
 seminar participants at Columbia, Harvard/MIT, UPenn, and Wisconsin are greatly appreciated. All errors are our own.
}
}
\date{August 2016}

\begin{document}

\maketitle

\begin{abstract}
The frequentist method of simulated minimum distance (SMD) is widely used
in economics to estimate complex models with an intractable likelihood.
In other disciplines, a Bayesian approach known as Approximate Bayesian Computation
(ABC) is far more popular. This paper connects these two seemingly related approaches
to likelihood-free estimation
by means of a Reverse  Sampler that
 uses both optimization and importance weighting to target the posterior distribution.
 Its hybrid features enable
 us to analyze an ABC estimate from the perspective of SMD. We show
that an ideal ABC estimate can be obtained as a weighted average of a sequence of
SMD modes, each being the minimizer of the deviations between the data and
the  model. This contrasts with the SMD, which is the mode of  the average deviations.
 Using stochastic expansions, we provide a general characterization
 of  frequentist estimators and those based on
 Bayesian computations including Laplace-type estimators. Their differences  are illustrated using  analytical examples
and a simulation study of the dynamic panel model.
\end{abstract}



\bibliographystyle{harvard}
\baselineskip=18.0pt
\thispagestyle{empty}
\setcounter{page}{0}
\newpage
\section{Introduction}
As knowledge accumulates, scientists and social scientists
 incorporate more  and more features into their models  to have a better
 representation of the data. The increased model complexity comes at a
 cost; the
conventional approach of estimating a model by writing down its
likelihood function is often not possible. Different disciplines have
developed different ways of handling models with an intractable
likelihood.
An approach popular amongst evolutionary biologists,
geneticists, ecologists,  psychologists and statisticians
is Approximate Bayesian Computation (ABC).  This work is largely unknown to economists who
mostly estimate complex models using frequentist
methods that we generically refer to as the method of Simulated Minimum Distance
(SMD),  and which include such estimators as  Simulated Method of Moments, Indirect
Inference, or Efficient Methods of Moments.\footnote{
Indirect Inference is due to \citet{gmr}, the Simulated Method
of moments is due to \citet{duffie-singleton},  and the
Efficient Method of Moments is due to \citet{gallant-tauchen-emm}.}

The  ABC and SMD  share the same
goal of estimating  parameters $\theta$ using auxiliary statistics $\hat\psi$ that are informative
about the data. An SMD estimator minimizes the $\mathsf L_2$ distance between $\hat\psi$ and  an average of  the
 auxiliary
statistics simulated under $\theta$, and  this distance can be made as close to zero as machine precision
permits.  An ABC estimator
evaluates the distance between $\hat\psi$
and the auxiliary statistics simulated for each  $\theta$ drawn from a proposal distribution.  The posterior mean is then
a weighted average of the draws that satisfy  a distance threshold of $\delta>0$.
There are many ABC algorithms, each differing according to
the choice of the distance metric, the weights,
 and sampling scheme. But  the algorithms
can only  approximate the desired posterior distribution because $\delta$ cannot
 be zero, or even too close to zero, in practice.

 While both  SMD and ABC use  simulations  to match $\psi(\theta)$
to  $\hat\psi$ (hence likelihood-free),
the relation between them  is not well understood beyond the fact that they
are asymptotically equivalent under some high level conditions.
To make progress, we focus on the MCMC-ABC algorithm due to \citet{mmpt-03}. The algorithm applies
uniform weights to those $\theta$ satisfying $\|\hat\psi-\psi(\theta)\|\le
\delta$ and zero otherwise.  Our main insight is that
 this  $\delta$  can  be made  very close to zero
if we combine optimization with Bayesian computations. In particular, the desired ABC posterior
 distribution can be targeted
using a `Reverse Sampler' (or RS for short) that applies importance
 weights to a sequence of SMD solutions.   Hence,  seen
 from the perspective
of the RS, the ideal MCMC-ABC estimate with $\delta=0$ is  a weighted  average of SMD modes. This offers
a useful contrast with the SMD estimate,
which is the mode of the
 average  deviations between the model and the data.
We then use stochastic expansions to study sources of variations in  the two
estimators in the case of exact identification.   The differences are illustrated
using simple analytical examples as well as
simulations of the dynamic panel model.


Optimization of
 models with a non-smooth  objective function is  challenging, even when
 the model is not complex.
The  Quasi-Bayes (LT) approach due
to  \citet{chernozhukov-hong} use Bayesian
computations to approximate  the  mode of a likelihood-free objective function.
Its validity rests on the Laplace
(asymptotic normal) approximation of the posterior distribution
 with the goal of valid asymptotic frequentist inference.
The simulation analog of the LT
(which we call SLT) further  uses simulations to approximate the intractable
 relation between the model and the data.
We show that both the LT and SLT can also be represented as  a weighted average
of modes with appropriately defined importance weights.

A central theme of our analysis is that the mean computed from many likelihood-free
posterior distributions can be seen as a weighted average of solutions to
frequentist objective functions.
Optimization  permits us to turn the focus from computational  to  analytical
aspects of the  posterior mean, and to
provide a bridge between the seemingly related approaches.
Although our optimization-based samplers  are not intended to compete  with
the many ABC algorithms that are available,
they can  be useful in situations when numerical
optimization of the auxiliary model is fast. This aspect is studied in our
companion paper  \citet{jjng-15} in which implementation of the RS
in the overidentified case is also considered. The RS is independently proposed
in \citet{meeds-welling} with emphasis on efficient and parallel implementations. Our
focus on the analytical properties complements their analysis.

The paper proceeds as follows. After laying out the preliminaries in
Section 2,  Section 3 presents the general idea behind ABC  and introduces
an optimization view of the ideal MCMC-ABC. Section 4 considers Quasi-Bayes estimators
 and interprets them from an optimization perspective.
Section 5 uses stochastic expansions to study the properties
 of the estimators.   Section 6 uses analytical examples and simulations to illustrate
 their differences. Throughout, we focus the discussion on
features  that distinguish the SMD from  ABC which are lesser known to economists.\footnote{
  The class of SMD estimators considered
 are well known  in the macro and finance literature and with
apologies, many references are  omitted. We also do not consider discrete choice
models; though the idea is conceptually similar,
the implementation requires different analytical tools.
 \citet{smith-palgrave} provides
a concise overview of these methods. The finite sample properties of
the estimators are studied in \citet{michaelides-ng}.  Readers are referred to the original paper
concerning the assumptions used.}

\section{Preliminaries}
 As a matter of notation, we use $L(\cdot)$ to denote the likelihood,
$p(\cdot)$ to denote posterior densities, $q(\cdot)$ for proposal densities,
and $\pi(\cdot) $ to denote prior
densities. A `hat'  denotes estimators that correspond to
 the mode and a `bar' is used for estimators that correspond to the posterior
 mean. We use
 $(s,S)$ and $(b,B)$ to denote the (specific, total number of)
draws in frequentist  and Bayesian type analyses respectively.
 A  superscript
$s$ denotes a specific draw and  a  subscript $S$  denotes the average over $S$ draws.
For a function $f(\theta)$,
we use $f_\theta(\theta_0)$ to denote $\frac{\partial}
{\partial \theta}f(\theta)$ evaluated at $\theta_0$,  $f_{\theta\theta_j}(\theta_0)$
to denote $\frac{\partial }{\partial \theta_j} f_\theta
(\theta)$ evaluated at $\theta_0$ and $f_{\theta,\theta_j,\theta_k}(\theta_0)$
to denote $\frac{\partial^2 }{\partial \theta_j \theta_k} f_\theta
(\theta)$ evaluated at $\theta_0$.

Throughout, we assume that the data
$\mathbf y=(y_1,\ldots,y_T)^\prime
$  are covariance stationary and can be represented by a parametric model
with probability measure $\mathcal P_\theta$ where $\theta \in \Theta\subset
\mymathcal R^K$.
The true value of $\theta$ is denoted by $\theta_0$. Unless otherwise stated, we write $\mathbb E[\cdot]$
for expectations  taken under $P_{\theta_0}$ instead of $\mathbb E_{\mathcal
P_{\theta_0}}[\cdot]$.
If the likelihood $L(\theta)=L(\theta|\mathbf y)$
 is tractable,   maximizing the log-likelihood $\ell
 (\theta)=\log L(\theta) $
with respect to $\theta$ gives
\[ \hat\theta_{ML}=\argmax_\theta \ell(\theta).\]


 Bayesian estimation  combines the likelihood with
a prior $\pi(\theta)$ to yield the posterior density
\begin{equation}
\label{eq:bc} p(\theta|\mathbf y)=\frac{L(\theta)\cdot \pi(\theta)}{\int_\Theta
L (\theta)\pi (\theta)
d\theta}.\end{equation}
For any  prior $\pi(\theta)$,  it is known that  $\hat\theta_
{ML}$ solves
  $\argmax_\theta \ell(\theta)=\lim_{\lambda\rightarrow \infty} \frac
  { \int_\Theta \theta\exp(\lambda \ell(\theta)) \pi(\theta
 )d\theta}{\int_\Theta \exp(\lambda \ell(\theta))\pi(\theta)d\theta}$.
That is, the maximum likelihood estimator
is a limit of the Bayes estimator
using   $\lambda\rightarrow\infty$ replications
 of the data $\mathbf y$.\footnote{See
 \citet[Corollary 5.11]{robert-casella},  \citet{jjp-07}.}
The parameter $\lambda$ is the cooling temperature
in simulated annealing, a stochastic optimizer
due to \cite{kirkpatrick-gellatt-vecchi} for handling problems with multiple
modes.


In the case of conjugate problems,  the posterior distribution has a parametric form which makes
it easy to  compute the posterior mean and other quantities of
interest.
For  non-conjugate problems, the method of Monte-Carlo Markov
Chain (MCMC) allows sampling from a Markov Chain
whose ergodic distribution is the target
posterior distribution $p(\theta|\mathbf y)$, and without the need to compute
the normalizing constant.
We use the  Metropolis-Hastings (MH) algorithm in subsequent
discussion. In classical Bayesian estimation  with proposal density $q(\cdot)$,
the acceptance ratio is
 \[\rho_{BC}(\theta^b,\theta^{b+1}) = \min \Big( \frac{L(\theta^{b+1})\pi(\theta^{b+1})q(\theta^b|\theta^{b+1}) }{L
 (\theta^b)\pi(\theta^b)q(\theta^{b+1}|\theta^b)},1 \Big).\]
When the posterior mode  $\hat\theta_{BC}=\argmax_\theta p(\theta|y)$ is difficult
to obtain,  the  posterior mean
\begin{eqnarray*}
\bar\theta_{BC}&=& \frac{1}{B}\sum_{b=1}^B \theta^b\approx \int_\Theta
\theta p(\theta|y)d\theta
  \end{eqnarray*}
is often the reported estimate,
where $\theta^b$ are draws  from the Markov Chain upon convergence.
Under quadratic loss, the posterior mean minimizes the posterior risk
$ Q(a)= \int_\Theta |\theta-a|^2 p(\theta|\mathbf y) d\theta$.


\subsection{Minimum Distance Estimators}
The method of generalized method of moments (GMM)
 is a likelihood-free frequentist estimator developed in \citet{hansen-82,hansen-singleton:82}.
For example, it allows for the estimation of $K$ parameters in a  dynamic model
 without explicitly solving the full model.
It is based on  a vector of $L\ge K$
moment conditions $ g_t
(\theta) $ whose expected value  is  zero at $\theta=\theta_0$, i.e. $\mathbb
E [g_t(\theta_0)]=0$.  Let $
\bar g(\theta)=\frac{1}{T}\sum_{t=1}^T g_t(\theta)$ be the sample analog
of $\mathbb E[g_t(\theta)]$.
The  estimator   is
\begin{eqnarray}
\label{eq:gmm}
\hat\theta_{GMM}&=&\argmin_
\theta  J(\theta), \quad\quad J(\theta)=\frac{T}{2}\cdot \bar g(\theta)^\prime W \bar g(\theta)
\end{eqnarray}
where $W$ is a $L\times L$ positive-definite weighting matrix.
Most estimators can be put in the GMM framework with suitable choice of $g_t$.
For example, when $g_t$ is the score of the likelihood,
the maximum likelihood estimator is obtained.

Let $\hat\psi\equiv \hat\psi(\mathbf
y(\theta_0))$  be $L$ auxiliary statistics with the property that $\sqrt{T}(\hat\psi -\psi(\theta_0))
\dconv \mathcal N(0,\Sigma)$. It is assumed that
 the mapping  $\psi(\theta)=\lim_{T\rightarrow\infty}
\mathbb E[\hat\psi(\theta)]$  is continuously differentiable in $\theta$
and locally injective at $\theta_0$.   \citet{gmr} refer to $\psi(\theta)$
as the {\em binding function}
  while  \citet{jiang-turnbull:94}
use the term {\em bridge function}.
 The minimum  distance estimator is a GMM estimator which specifies
\[ \bar g(\theta)=\hat \psi- \psi(\theta),\]
with efficient weighting matrix $W=\Sigma^{-1}$. Classical MD estimation assumes that the binding function $\psi(\theta)$
has a closed form expression so that in the exactly identified case, one can
solve for $\theta$ by inverting $\bar g(\theta)$.

\subsection{SMD Estimators}


Simulation estimation is
useful when
the asymptotic binding function  $\psi(\theta_0)$
 is not analytically tractable but can be easily evaluated on simulated data.
The first use of this approach in economics appears to be
due to \citet{smith-93}.
The simulated analog of MD, which we will call SMD,
 minimizes
the weighted difference between the auxiliary statistics evaluated
at the observed and simulated data:
\begin{eqnarray*}\hat\theta_{SMD}&=&\argmin_
\theta J_S(\theta)=\argmin_\theta \bar g_S^{\prime}(\theta) W \bar g_S(\theta).
\end{eqnarray*}
where
\[   \bar g_S(\theta) =  \hat\psi -\frac{1}{S}\sum_{s=1}^S \hat
\psi^s(\mathbf y^s(\theta)),
\]
$\mathbf y^s(\theta) \equiv \mathbf y^s(\varepsilon^s,\theta)$ are data simulated under $\theta$ with errors $\varepsilon^s$
drawn from an assumed distribution $F_\varepsilon$, and $\hat\psi^s(\theta)\equiv
\hat\psi^s (\mathbf y^s
(\varepsilon^s,\theta))$ are the auxiliary statistics computed using $\mathbf
y^s(\theta)$. Of course, $\bar g_S(\theta)$ is also the average over $S$ deviations
between $\hat\psi$ and $\hat\psi^s(\mathbf y^s(\theta))$. To
simplify notation, we will write $\mathbf y^s$ and $\hat\psi^s(\theta)$
when the context is clear.
As in MD estimation, the  auxiliary statistics $\psi(\theta)$ should
`smoothly embed' the properties of the data in the terminology of \citet{gallant-tauchen-emm}.
But SMD estimators  replace the asymptotic binding function  $\psi(\theta_0)=\lim_{T\rightarrow\infty}
\mathbb E [\hat\psi
(\theta_0)]$ by a finite sample analog using Monte-Carlo
simulations.   While the SMD is motivated with the estimation of complex models
in mind, \citet{grt-99} show that simulation estimation has a
bias reduction effect like the bootstrap. Hence in the econometrics literature,
SMD estimators are used even when the likelihood is tractable, as in \citet{gpy}.

The steps for implementing the SMD are as follows:
\begin{itemize}
\item[0] For $s=1,\ldots, S$, draw
$\varepsilon^s=(\varepsilon^s_1,\dots,\varepsilon^s_T)^\prime$ from
$F_\varepsilon$. These are
  innovations to the structural model that will be held fixed during
iterations.

\item[1] Given $\theta$, repeat  for $s=1,\ldots S$:
\begin{enumerate}
\item[a] Use $(\varepsilon^s$, $\theta)$ and the model to
simulate data $\mathbf y^s=(y_1^s,\ldots, y_T^s
)^\prime$.
\item[b] Compute the auxiliary statistics $\hat\psi^s(\theta)$ using simulated data
$\mathbf y^s$.
\end{enumerate}
\item[2]  Compute: $ \bar g_S(\theta)=\hat\psi(\mathbf y)-\frac{1}{S}\sum_{
s=1}^S \hat\psi^s(\theta)$.
Minimize $J_S(\theta)= \bar g_S(\theta)^\prime
W \bar g_S(\theta)$.
\end{itemize}
The SMD is the $\theta$ that makes $J_S(\theta)$
smaller than the tolerance specified for the numerical optimizer.
In the exactly identified case, the tolerance can be made as small as machine
precision permits.
When $\hat\psi$ is a vector of unconditional moments, the SMM estimator of
\citet{duffie-singleton} is obtained. When $\hat\psi$ are parameters
of an auxiliary model, we have the `indirect inference'
estimator of \citet{gmr}. These are Wald-test
based SMD estimators in the terminology of \citet{smith-palgrave}. When $\hat\psi$ is the score function
associated with the likelihood of the auxiliary model, we have the EMM
estimator of \citet{gallant-tauchen-emm}, which can also be thought of as an LM-test
based SMD. If $\hat\psi$ is the
likelihood of the auxiliary model, $J_S(\theta)$ can be interpreted as a likelihood
ratio and we have a LR-test based SMD.
 \citet{g-monfort-simulation} provide a framework that unifies
these three approaches to SMD estimation.  \citet{nickl-potscher} show that an SMD
based on non-parametrically  estimated  auxiliary statistics
 can have asymptotic variance equal to the Cramer-Rao bound
  if the tuning parameters are optimally chosen.

The Wald, LM, and LR based SMD estimators  minimize a
weighted $\mathsf L_2$ distance
between the data and the model as summarized by  auxiliary statistics.
\citet{creel-kristensen-il} consider a class of estimators that minimize
the Kullback-Leibler distance between the model and the data.\footnote{
In the sequel,
we take  the more conventional $\mathsf L_2$ definition of SMD as given above.}
Within this class, their  MIL estimator maximizes an `indirect likelihood', defined
as the likelihood of the auxiliary statistics.
Their BIL estimator  uses Bayesian
computations to approximate the mode of the indirect likelihood. In practice, the indirect likelihood  is unknown. Estimating it by kernel
smoothing of the simulated statistics, the  SBIL estimator
 combines Bayesian computations with non-parametric
estimation.  \citet{gao-hong}
show that  using local linear regressions instead of kernel estimation can reduce the
variance and the bias.  Using non-parametric
estimation in ABC has previously been considered in \citet{beaumont-zhang-balding}.
 \citet{cghk:16}  show that not only can such an ABC implementation
bypass MCMC altogether, it can provide asymptotically valid frequentist inference. Bounds for the number of simulations
that achieve the parametric rate of convergence and asymptotic normality are derived.


\section{Approximate Bayesian Computations}

The ABC literature often credits
Donald Rubin to be the first to  consider the possibility of
 estimating the posterior distribution  when the likelihood  is
 intractable.  \citet{diggle-gratton-84}
propose to approximate the likelihood by simulating the model at each point on a
parameter grid and appear to be the first implementation of
simulation estimation for
models with intractable likelihoods. Subsequent developments
adapted the idea to conduct posterior inference, giving the prior  an explicit role.
The first ABC algorithm was implemented  by \citet{tbfd} and \citet{pspf:99} to study
population genetics. Their  Accept/Reject algorithm is as follows: (i) draw $\theta^b$ from the prior
distribution $\pi(\theta)$, (ii) simulate data using the model  under $\theta^b$
 (iii)  accept $\theta^b$ if the auxiliary
statistics computed using the simulated data  are close
 to $\hat\psi$.   As in the SMD literature, the auxiliary statistics can be
parameters of a regression or  unconditional sample moments. \citet{heggland-frigessi},
\citet{drovandi-pettitt-faddy,drovandi-15}
use simulated auxiliary statistics.

 Since simulating from a non-informative prior distribution is inefficient,
 subsequent work suggests to replace the
rejection sampler by one that takes into account the features of the posterior distribution.
The likelihood of the full dataset $L(y|\theta)$ is intractable, as is the likelihood of the finite dimensional statistic $L(\hat \psi|\theta)$. However, the latter can be consistently estimated using simulations.
The general idea is to set as a target the intractable posterior density
\[p^*_{ABC}(\theta|\hat \psi) \propto \pi(\theta)L
(\hat \psi|\theta)\]
and approximate it   using Monte-Carlo methods.
 Some algorithms are  motivated from the perspective of non-parametric density estimation, while others aim
to improve  properties of the Markov chain.\footnote
{
Recent surveys on ABC can be found in  \citet{mprr-12},
\citet{blum-nunes-prangle-sisson} among others. See \citet{drovandi-15,drovandi-pettitt-faddy}
for differences amongst ABC estimators.} The main idea is, however, using
data augmentation to consider the joint density $p_{ABC}(\theta,x|\hat\psi)\propto
L(\hat\psi|x,\theta)L(x|\theta)\pi(\theta)$,  putting more weight on the draws with $x$ close to $\hat\psi$.
When $x=\hat\psi$, $L(\hat\psi|\hat\psi,\theta)$ is a constant,  $p_{ABC}
(\theta,\hat\psi|\hat\psi)\
\propto L(\hat\psi|\theta)\pi(\theta)$, and the target posterior is recovered.
If $\hat\psi$ are sufficient
 statistics, one recovers the posterior distribution associated with the intractable
 likelihood $L(\theta|y)$, not just an approximation.

To better understand the ABC idea and its implementation, we will write $\mathbf y^{b}$ instead of
$\mathbf y^{b}
(\varepsilon^{b},\theta^{b})$ and $\hat \psi^{b}$ instead of $\hat\psi^{b}(\mathbf y^{b}
(\varepsilon^{b},\theta^{b}))$ to
simplify notation.
Let $\mathbb K_\delta(\hat \psi^b,\hat \psi|\theta)\geq 0$ be a kernel function that
weighs  deviations between $\hat\psi$ and $\hat\psi^b$ over a window of width
$\delta$. Suppose we
keep only the draws that satisfy
$\hat \psi^b=\hat \psi$ and hence $\delta=0$. Note that $\mathbb K_0(\hat\psi^b,\hat\psi|\theta)=1$
if $\hat\psi=\hat\psi^b$ for any choice of the kernel function.
Once the likelihood of interest
\[ L(\hat \psi|\theta) = \int L(x|\theta)\mathbb K_0(x,\hat \psi|\theta)dx \]
is available, moments and quantiles can be computed.
In particular, for any measurable function $\varphi$ whose expectation exists, we have:
\begin{align*}
  \mathbb{E} \left[\varphi(\theta)|\hat \psi=\hat\psi^b \right] &= \frac{\int_\Theta
  \varphi(\theta^b) \pi(\theta)L(\hat \psi | \theta^b)d\theta^b}{\int_\Theta \pi
  (\theta^b)L(\hat \psi | \theta^b)d\theta^b}
  = \frac{\int_\Theta \int \varphi(\theta^b) \pi(\theta^b) L(x|\theta^b)\mathbb
  K_0(x,\hat \psi|\theta^b)dxd\theta^b}{\int_\Theta \int \pi(\theta^b) L(x|\theta^b)\mathbb K_0(x,\hat \psi|\theta^b)dxd\theta^b}.
\end{align*}
Since $\hat \psi^b|\theta^b \sim L(\cdot|\theta^b)$, the expectation can be approximated
by averaging over draws from $L(\cdot|\hat\theta^b)$. More generally,  draws
can be taken from an importance density $q(\cdot)$. In particular,
\[   \hat{\mathbb E}\left[\varphi
(\theta)|\hat\psi=\hat\psi^b\right]=\frac{\sum_{b=1}^B
\varphi(\theta^b) \mathbb K_0(\hat\psi^b,\hat \psi|\theta^b)\frac{\pi(\theta^b)}{q
(\theta^b)}}{\sum_{b=1}^B \mathbb K_0(\hat\psi^b,\hat \psi|\theta^b)\frac{\pi(\theta^b)}{q(\theta^b)}}.\]
The importance weights are then
 \[w_0^b \propto \mathbb K_0(\hat\psi^b,\hat \psi|\theta^b)\frac{\pi(\theta^b)}{q
(\theta^b)}.\]
By a law of large numbers,
$\hat{\mathbb{E}} \left[ \varphi(\theta)|\hat \psi\right]\rightarrow \mathbb
{E} \left[ \varphi(\theta)|\hat \psi\right]$ as $B\rightarrow\infty$.

There is, however, a caveat. When $\hat \psi$ has continuous support,  $\hat \psi^b
 = \hat \psi$ is an event of measure zero. Replacing
 $\mathbb K_0$ with $\mathbb K_\delta$ where $\delta$ is close
 to zero  yields the approximation:
 \begin{eqnarray*} \mathbb
 {E}\left[\varphi(\theta)|\hat \psi=\hat\psi^b\right] &\approx &\frac{\int_\Theta \int \varphi
 (\theta^b) \pi(\theta^b) L(x|\theta^b)\mathbb K_\delta(x,\hat \psi|\theta^b)dxd\theta^b}{\int_\Theta
 \int \pi(\theta^b) L(x|\theta^b)\mathbb K_\delta(x,\hat \psi|\theta^b)dxd\theta^b}.
 \end{eqnarray*}
 Since $\mathbb K_\delta(\cdot)$ is a kernel function,  consistency of the
 non-parametric estimator
 for  the conditional expectation of $\varphi(\theta)$
follows from, for example, \citet{pagan-ullah}.
 This is the approach considered in \citet{beaumont-zhang-balding}, \citet
 {creel-kristensen-il} and  \citet{gao-hong}.
The case of a  rectangular  kernel  $\mathbb K_\delta
(\hat \psi,\hat \psi^b) = \mathbbm I_{\|\hat \psi-\hat\psi^b\|\leq \delta}$
corresponds to
the ABC algorithm proposed in \citet{mmpt-03}.  This is the first
 ABC algorithm that exploits MCMC  sampling. Hence we  refer to it as
 MCMC-ABC.
Our analysis to follow is based  on this algorithm. Accordingly,
 we now explore it
  in more detail.

\paragraph{Algorithm MCMC-ABC}
Let $q
(\cdot)$ be the proposal distribution. For $b=1,\ldots, B$ with $\theta^0$ given,
\begin{enumerate}
\item[1] Generate $\theta^{b+1} \sim q(\theta^{b+1}|\theta^b)$.
\item[2] Draw  $ \varepsilon^{b+1}$ from $F_\varepsilon$ and  simulate data $\mathbf
 y^{b+1}$. Compute
 $\hat{\psi}^{b+1}$.
\item[3] Accept $\theta^{b+1}$  with probability $ \rho_{\text{ABC}}
(\theta^b,\theta^{b+1}) $ and set it equal to
$\theta^b$ with probability $ 1-\rho_{\text{ABC}}(\theta^b,\theta^{b+1})$ where
 \begin{equation}
 \label{eq:rhoABC-simple}
 \rho_{\text{ABC}}(\theta^b,\theta^{b+1}) = \min \Big( \mathbbm I_{\|\hat\psi-\hat\psi^{b+1}\|\leq \delta} \frac{\pi(\theta^{b+1})q(\theta^b|\theta^{b+1})}{\pi(\theta^b)q(\theta^{b+1}|\theta^b)}
,1 \Big  ).\end{equation} \end{enumerate}
As with all ABC algorithms, the success of the MCMC-ABC lies in augmenting the posterior with simulated
data $\hat \psi^b$, i.e. $p^*_{ABC}(\theta^b,\hat\psi^b|\hat\psi)\propto L
(\hat\psi|\theta^b,\hat\psi^b)L
(\hat\psi^b|\theta^b)\pi(\theta^b)$.
The  joint posterior distribution that the MCMC-ABC would
like to target is
\[ p^0_{\text{ABC}}\left( \theta^b, \hat \psi^b | \hat \psi \right) \propto
\pi(\theta^b)L(\hat \psi^b|\theta^b)\mathbbm I_{\|\hat \psi^b-\hat \psi\|
= 0} \]
since integrating out $\varepsilon^b$ would yield $p^*_{ABC}(\theta|\hat\psi)$.
But it would not be possible to generate draws such that $\|\hat\psi^b-\hat\psi\|$
equals zero exactly.
Hence as a compromise, the MCMC-ABC algorithm allows $\delta>0$ and targets
\[ p^\delta_{\text{ABC}}\left( \theta^b, \hat \psi^b | \hat \psi \right) \propto
\pi(\theta^b)L(\hat \psi^b|\theta^b)\mathbbm I_{\|\hat \psi^b-\hat \psi\|
\leq \delta}. \]
The adequacy of $p_{ABC}^\delta$ as an approximation of $p^0_{ABC}$ is a function
of the tuning parameter $\delta$.

To understand why this algorithm works, we follow the argument in \citet{sisson-fan}.
If the initial draw $\theta^1$ satisfies $\|\hat \psi - \hat \psi^1 \| \leq
\delta$, then   all subsequent $b>1$ draws are such that
  $ \mathbbm I_{\| \hat \psi^b - \hat \psi \| \leq \delta } =1$ by construction.
Furthermore,
since we draw $\theta^{b+1}$ and then independently simulate data $\hat \psi^{b+1}$, the proposal distribution becomes
$ q(\theta^{b+1},\hat\psi^{b+1}|\theta^b) = q(\theta^{b+1}|\theta^b)L(\hat\psi^
{b+1}|\theta^{b+1}). $
The two observations together imply that
  \begin{align*}
     \mathbbm I_{\|\hat\psi-\hat\psi^{b+1}\|\leq \delta} \frac{\pi(\theta^{b+1})q(\theta^b|\theta^{b+1})}{\pi(\theta^b)q(\theta^{b+1}|\theta^b)} &=
      \frac{\mathbbm I_{\|\hat\psi-\hat\psi^{b+1}\|\leq \delta}}{\mathbbm I_{\|\hat\psi-\hat\psi^{b}\|\leq \delta}} \frac{\pi(\theta^{b+1})q(\theta^b|\theta^{b+1})}{\pi(\theta^b)q(\theta^{b+1}|\theta^b)}\frac{L(\hat\psi^{b+1}|\theta^{b+1})}{L(\hat\psi^{b}|\theta^{b})}\frac{L(\hat\psi^{b}|\theta^{b})}{L(\hat\psi^{b+1}|\theta^{b+1})} \\
      &= \frac{\mathbbm I_{\|\hat\psi-\hat\psi^{b+1}\|\leq \delta}}{\mathbbm I_{\|\hat\psi-\hat\psi^{b}\|\leq \delta}} \frac{\pi(\theta^{b+1})L(\hat\psi^{b+1}|\theta^{b+1})}{\pi(\theta^b)L(\hat\psi^{b}|\theta^{b})}\frac{q(\theta^b|\theta^{b+1})L(\hat\psi^{b}|\theta^{b})}{q(\theta^{b+1}|\theta^b)L(\hat\psi^{b+1}|\theta^{b+1})} \\
      &=\frac{p^\delta_{\text{ABC}}\left( \theta^{b+1}, \hat \psi^{b+1} |
      \hat \psi \right)}{p^\delta_{\text{ABC}}\left( \theta^b, \hat \psi^b | \hat \psi \right)}\frac{q(\theta^b,\hat \psi^b|\theta^{b+1})}{q(\theta^{b+1},\hat \psi^{b+1}|\theta^b)}.
  \end{align*}
The last equality shows that the acceptance ratio is in fact the ratio of
two ABC posteriors  times the ratio of the proposal distribution. Hence the
MCMC-ABC effectively targets the joint posterior distribution $p_{ABC}^\delta$.

\subsection{The Reverse Sampler}
Thus far, we have seen that the SMD estimator is the $\theta$ that makes
$\|\hat\psi-\frac{1}{S}\sum_{s=1}^S\hat\psi^s(\theta)\|$  no larger than
the tolerance of the numerical optimizer. We have also seen that the
feasible MCMC-ABC
accepts  draws $\theta^b$ satisfying $\|\hat\psi-\hat\psi^b(\theta^b)\|\leq
\delta $ with $\delta>0$. To view the  MCMC-ABC from a different perspective,
suppose that setting $\delta=0$ was possible. Then each accepted draw $\theta^b$ would satisfy:
\[ \hat \psi^b(\theta^b)=\hat \psi. \]
For fixed $\varepsilon^b$ and assuming that the mapping $\hat \psi^b : \theta
\rightarrow \hat \psi^b(\theta)$ is continuously differentiable and one-to-one, the above statement is equivalent to:
\[ \theta^b = \text{argmin}_\theta \left( \hat \psi^b(\theta)-\hat \psi\right)^\prime \left( \hat \psi^b(\theta)-\hat \psi\right). \]
Hence each accepted $\theta^b$ is the solution to a SMD problem with $S=1$.
Next, suppose that instead of drawing $\theta^b$ from a proposal distribution, we draw $\varepsilon^b$ and solve for $\theta^b$ as above.
Since  the mapping $\hat \psi^b$ is invertible by
assumption,  a change of variable yields the relation between the distribution
of $\hat\psi^b$ and $\theta^b$. In particular, the joint density, say  $h(\theta^b,\varepsilon^b)$,
is related to the joint density $L(\hat\psi^b(\theta^b),\varepsilon^b)$
via the determinant of the Jacobian $|\hat \psi^b_\theta
 (\theta^b)|$ as follows:
 \[ h(\theta^b,\varepsilon^b|\hat \psi) = |\hat \psi^b_\theta
 (\theta^b)|L(\hat \psi^b(\theta^b), \varepsilon^b|\hat \psi).
 \]
 Multiplying the quantity on the right-hand-side  by $w^b(\theta^b)=\pi(\theta^b)|\hat \psi^b_\theta
 (\theta^b)|^{-1}$ yields $\pi(\theta^b)L(\hat\psi,\varepsilon^b|\theta^b)$
 since $\hat\psi^b(\theta^b)=\hat\psi$ and the mapping from
 $\theta^b$ to $\psi^b(\theta^b)$ is one-to-one.
 This suggests that if we  solve the SMD problem $B$ times each with
 $S=1$, re-weighting each of the $B$ solutions by $w^b(\theta^b)$ would give the target the joint
 posterior $p_{ABC}^*(\theta|\hat\psi)$ after integrating out $\varepsilon^b$.


  \paragraph{Algorithm RS }
\begin{itemize}
\item[1]For $b=1,\ldots, B$ and a given $\theta$,
\begin{itemize}
\item[i] Draw $\varepsilon^b $ from $F_\varepsilon$ and simulate
data $ \mathbf y^b$ using $\theta$. Compute $\hat\psi^b(\theta)$ from $\mathbf
y^b$.
\item[ii] Let  $\theta^b = \argmin_{\theta}  J_1^b(\theta)$, $J_1^b(\theta
)=(\hat\psi-\hat\psi^b(\theta))^\prime W (\hat \psi-\hat\psi^b(\theta
))$.
\item[iii] Compute the Jacobian $\hat\psi_\theta^b(\theta^b)$ and its determinant
$|\hat\psi_\theta^b(\theta^b)|$. Let $w^b(\theta^b)=\pi(\theta^b)|\hat{\psi}^b_\theta(\theta^b)|^{-1}$. \end{itemize}
\item[2] Compute the posterior mean
 $ \bar\theta_{RS }=\sum_{b=1}^B \bar w^b(\theta^b)
\theta^b $ where $\bar w^b(\theta^b)=\frac{w^b(\theta^b)}{\sum_{c =1}^B w^c
 (\theta^{c})}$.
\end{itemize}
The RS  has the optimization aspect of SMD as well as the sampling aspect
of the MCMC-ABC.
We call the RS the reverse sampler for two reasons. First, typical Bayesian estimation starts with an evaluation of
the prior probabilities. The RS   terminates with the evaluation of the prior.
Furthermore, we use the SMD estimates to reverse engineer the posterior distribution.

Consistency of each RS solution (i.e. $\theta^b$)  is built on the fact that
the SMD is consistent  even with $S=1$. The RS estimate is thus an average of a sequence
of SMD modes. In contrast, the SMD  is the mode of an objective function defined
from a weighted average of the simulated auxiliary statistics.
Optimization
effectively allows $\delta$  to be as close to zero as machine precision permits.
This puts the joint posterior distribution as close to
 the infeasible target  as possible,
 but has the consequence of shifting  the distribution
from $(\mathbf y^b,\hat\psi^b)$ to $(\mathbf y^b,\theta^b)$. Hence  a change of
variable is required. The importance weight depends on the Jacobian matrix, making
the RS an optimization based importance sampler.

\begin{lemma}
\label{prop:prop1} Suppose that $\psi: \theta\rightarrow \hat\psi^b(\theta)$ is one-to-one and
$\psi^b_\theta(\theta)$ has full column rank.
The  posterior distribution produced by the reverse sampler converges
to the infeasible  posterior distribution $p^*_{ABC}(\theta|\hat\psi)$ as $B\rightarrow\infty$.

\end{lemma}
The proof is given in \citet{jjng-15}.
By convergence, we mean that for any measurable function $\varphi(\theta)$
such that the expectation exists, a law of large numbers implies that \newline
$ \sum_{b=1}^B \bar w^b(\theta^b) \varphi(\theta^b)
 \asconv \mathbb E_{p^*(\theta|\hat\psi)}(\varphi(\theta))$. In general, $\bar
 w^b(\theta^{b})\ne \frac1B$. The
 RS draws and moments can be interpreted
as if they were taken from  $p^*_{\text
 {ABC}}$, the posterior distribution
had the likelihood $p(\hat\psi|\theta)$
been available.

That the draws of the MCMC-ABC at $\delta=0$ can be seen from an optimization perspective
allows us to subsequently use
the RS as a  conceptual framework to understand the differences
between the ideal  MCMC-ABC and SMD. It should be noted that
the RS is not the same as the MCMC-ABC or any ABC estimator implemented with $\delta>0$ as
they necessarily have an acceptance rate strictly less than one. Indeed, a challenge of many ABC implementations is
the low acceptance rate.
 The RS draws are always accepted and can be useful in situations when numerical
optimization of the auxiliary model is easy.
Properties of the RS  are further analyzed in \citet{jjng-15}. \citet {meeds-welling}
 independently propose an ABC sampling algorithm similar to the RS. Their focus is
 on ways to   implement it efficiently using  embarrassingly parallel methods.

\section{Quasi-Bayes Estimators}

The GMM objective function $J(\theta)$ defined in (\ref{eq:gmm}) is not a proper
density. Noting that $\exp(-J(\theta))$ is the kernel of the Gaussian density,
 \citet{
jiang-turnbull:94} define an {\em indirect likelihood} (distinct from the
one defined in \citet{creel-kristensen-il}) as
\[ L_{IND}(\theta|\hat\psi) \equiv \frac{1}{\sqrt{2\pi}} |\Sigma|^{-1}
\exp( - J(\theta)).\]
Associated with the indirect likelihood is the indirect score,
indirect Hessian, and a generalized information matrix equality, just like a conventional likelihood.
 Though the indirect likelihood is not a proper density,
its maximizer  has properties
analogous to the maximum likelihood estimator provided by $\mathbb E[g_t
(\theta_0)]=0$.


In \citet{chernozhukov-hong}, the authors observe that
 extremum estimators can be difficult to compute if the objective function is highly
non-convex, especially when the dimension of the parameter space is large.
 These difficulties can be alleviated
by using Bayesian computational tools, but this is not possible when the objective
function is not a likelihood.
\citet{chernozhukov-hong}  take
 an  exponential  of $-J(\theta)$, as
 in \citet{jiang-turnbull:94},
but then combine  $\exp(-J(\theta))$ with a prior density $\pi(\theta)$ to produce a quasi-posterior density.
Chernozhukov and Hong
initially termed their estimator `Quasi-Bayes' because $\exp(-J(\theta))$ is not
a standard likelihood.  They
settled on the term `Laplace-type estimator' (LT),
so-called because Laplace suggested to  approximate
 a smooth pdf with a well defined
peak by a normal density, see \citet{tierney-kadane:86}.
If $\pi(\theta)$  is strictly
 positive and continuous over a compact parameter space $\Theta$, the `quasi-posterior' LT distribution
\begin{equation}
\label{eq:lt} p_{LT}(\theta|\mathbf y)=\frac{\exp(-J(\theta))\pi(\theta)}
{\int_\Theta
\exp (-J (\theta)\pi (\theta))d\theta
} \propto \exp(-J(\theta))\pi(\theta)
\end{equation}
 is  proper. The LT posterior mean is thus well-defined even when the prior may not be proper. As discussed in \citet{chernozhukov-hong},
one can think of the LT under a flat prior as using simulated annealing to maximize $\exp(-J(\theta))$
and setting the cooling parameter $\tau$ to 1.
 Frequentist inference is asymptotically
valid because as the sample size increases, the prior is dominated by the
pseudo likelihood which, by the Laplace approximation, is asymptotically normal.\footnote{
For loss function $d(\cdot)$,
 the LT estimator is
 $ \hat\theta_{LT}(\vartheta)=\argmin_\theta
 \int_\Theta d(\theta-\vartheta) p_{LT}(\theta|\mathbf y)d\theta.
 $
If $d(\cdot)$ is  quadratic,  the posterior mean  minimizes quasi-posterior risk.}

In practice, the LT posterior distribution is targeted using MCMC methods.
Upon replacing   the likelihood $L(\theta)$
by $\exp(-J(\theta))$,
the  MH acceptance probability is
 \[\rho_{LT}(\theta^b,\vartheta) = \min \Big( \frac{\exp(-J(\vartheta))\pi(\vartheta)q(\theta^b|\vartheta)}
 {\exp(
 -J(\theta^
b))\pi(\theta^b)q(\vartheta|\theta^b)},1 \Big).\]
The quasi-posterior mean
  is
  $ \bar \theta_{LT}= \frac{1}{B} \sum_{b=1}^B \theta^b$ where each $\theta^b$  is
a draw from $p_{LT}(\theta|\mathbf y)$.
Chernozhukov and Hong suggest to exploit the fact that
  the quasi-posterior mean is much easier to compute than the mode and that,
  under regularity conditions,   the two are first order equivalent. In practice,
  the weighting matrix can be based on some preliminary estimate of $\theta$, or
estimated simultaneously with $\theta$. In exactly identified models, it
is well known that the MD estimates do not depend on the choice of $W$. This
continues to be the case for the LT posterior mode $\hat\theta_{LT}$. However,
the posterior mean is affected by the choice of the weighting matrix even
in the just-identified case.\footnote{\cite{kormiltsina-nekipelov:14}
suggests to scale  the objective function to improve  coverage of the confidence
intervals.}

The LT estimator is built  on the validity of the asymptotic normal approximation
in the second order expansion of the objective function.    \cite{nekipolov-kormilitsina:15}
show that in small samples, this approximation can be poor so that
the LT posterior mean may differ significantly from the extremum estimate
that it is meant to approximate. 
To see the problem in a different light,  we again take an  optimization view.
Specifically,   the asymptotic distribution
$\sqrt{T}(\hat\psi(\theta_0)-\psi(\theta_0))\dconv
\mathcal N (0,\Sigma (\theta_0))\equiv \mathbb A_\infty(\theta_0)$
 suggests to use
 \[\hat\psi^b(\theta) \approx \psi(\theta) +\frac{\mathbb A^b_\infty(\theta_0)}{\sqrt{T}}\] where
$\mathbb A^b_\infty(\theta_0) \sim \mathcal N (0,\hat\Sigma(\theta))$.
Given a draw of $\mathbb A^b_\infty$, there will exist a
$ \theta^b$ such that $ (\hat\psi^b(\theta)-\hat\psi)^\prime W (\hat\psi^b(\theta)-\hat\psi
)$
is minimized.
In the exactly identified case, this discrepancy can be driven to zero up to
machine precision. Hence we can define
\[ \theta^b= \argmin_\theta \|\hat\psi^b(\theta)-\hat\psi\|.\]
Arguments analogous to the RS  suggest the following
will produce draws of $\theta$ from $p_{LT}(\theta|\mathbf y)$.
\begin{enumerate}
\item[1] For $b=1,\ldots $B:
\begin{itemize}
  \item[i] Draw $\mathbb A^b_\infty(\theta_0)$ and  define $\hat \psi^b(\theta)=\psi(\theta)+\frac{\mathbb A^b_\infty(\theta)}{\sqrt{T}}$.
  \item[ii] Solve for $\theta^b$ such that $\hat \psi^b(\theta^b)=\hat \psi$ (up to machine precision).
  \item[iii] Compute   $ w^b(\theta^b) = |\hat \psi_\theta^b(\theta^b)|^{-1}\pi(\theta^b)$.
  \end{itemize}
  \item[2] Compute $\bar \theta_{LT} = \sum \bar w^b(\theta^b)\theta^b$, where $\bar w^b=\frac{w^b
  (\theta^b)}{\sum_{c=1}^B w^c(\theta^c)}$.
\end{enumerate}
Seen from an optimization perspective, the LT is a weighted average of MD
modes  with the determinant of the Jacobian matrix as importance weight,
similar to the RS.
It differs from the RS in that  the Jacobian here is computed from
 the asymptotic binding function $\psi(\theta)$,
and  the draws are based on the asymptotic normality of $\hat\psi$.
 As such,  simulation of the structural model is not required.


\subsection{The SLT}

When $\psi(\theta
)$ is not analytically tractable,
a natural modification
is to approximate it  by simulations  as in the SMD.  This
is the approach taken in \citet{lise-meghir-robin}.
 We refer to this estimator
 as the Simulated Laplace-type estimator, or SLT.  The steps are as follows:
\begin{itemize}
  \item[0]   Draw structural innovations
  $ \varepsilon^{s}=( \varepsilon ^{s}_1,\ldots, \varepsilon
  ^{s}_T)^\prime$ from $F_\varepsilon$. These are held fixed across iterations.
\item[1] For $b=1,\ldots, B$, draw $\vartheta$ from $q(\vartheta|\theta^b)$.
\begin{itemize}
\item[i.]  For $s=1,\ldots S$:
 use $(\vartheta, \varepsilon^{s})$ and the model to simulate data $\mathbf
 y^s=(\mathbf y_1^s,\ldots,\mathbf y_T^s)^\prime$.
Compute $\hat\psi^s(\vartheta)$ using $\mathbf y^s$.
\item[ii.] Form  $J_S(\vartheta)=\bar g_S(\vartheta)^\prime W \bar g_S(\vartheta)$,
where $\bar g_S(\vartheta)=\hat\psi(\mathbf y)-\frac{1}{S}\sum_{s=1}^S
\hat\psi^s(\vartheta)$.
\item[iii.] Set $\theta^{b+1} =\vartheta$  with probability $\rho_{SLT}(\theta^b,\vartheta)$,
else reset $\vartheta$ to $\theta^b$ with probability $1-\rho_{SLT}$
where the acceptance probability is:
\[\rho_{SLT}(\theta^b,\vartheta) = \min \Big( \frac{\exp(-J_S(\vartheta))\pi(\vartheta)q(\theta^b|\vartheta)}
{\exp(
-J_S(\theta^
b))\pi(\theta^b)q(\vartheta|\theta^b)},1 \Big).\]
\end{itemize}
\item[2] Compute $\bar\theta_{SLT}^b=\frac{1}{B} \sum_{b=1}^B \theta^b$.
\end{itemize}
 The  SLT algorithm has two loops, one using $S$ simulations for each $b$ to approximate the
 asymptotic binding function, and one using $B$ draws to approximate
 the `quasi-posterior' SLT distribution
\begin{eqnarray}
\label{eq:slt} p_{SLT}(\theta|\mathbf y,\varepsilon^1,\ldots,\varepsilon^S)&=&
\frac{\exp(-J_S(\theta))\pi(\theta)}
{\int_\Theta
\exp(-J_S(\theta))\pi(\theta)d\theta
} \propto \exp(-J_S(\theta))\pi(\theta)
\end{eqnarray}

The above SLT algorithm has features of SMD, ABC, and LT, it also requires simulations of the full model.
As a referee pointed out, though the SLT resembles the ABC algorithm when used  with
a Gaussian kernel,  $\exp (-J_S(\theta))$ is  not
a proper density, and $p_{SLT}(\theta|\mathbf y,\varepsilon^1,\ldots,\varepsilon^S)$
is not a conventional likelihood-based posterior distribution. While the SLT targets the pseudo likelihood,
   ABC algorithms target the proper but intractable likelihood. Furthermore,
   the asymptotic distribution of $\hat\psi$ is known from a frequentist perspective.
   In ABC estimation, lack of knowledge of the likelihood of $\hat\psi$    motivates the
   Bayesian computation.

The optimization implementation of SLT presents a clear contrast with
 the ABC.
\begin{enumerate}
\item[1]  Given $ \varepsilon^{s}=( \varepsilon ^{s}_1,\ldots, \varepsilon
  ^{s}_T)^\prime$ for $s=1,\ldots S$, repeat for $b=1,\ldots B$:
\begin{enumerate}
  \item[i] Draw $\hat \psi^b(\theta) = \frac{1}{S}\sum_{s=1}^S  \hat \psi^s(\theta) + \frac{\mathbb A_\infty^b(\theta)}{\sqrt{T}}$.
  \item[ii] Solve for $\theta^b$ such that $\hat \psi^b(\theta^b)=\hat \psi$ (up to machine precision).
  \item[iii] Compute   $ w^b(\theta^b) = |\hat \psi_\theta^b(\theta^b)|^{-1}\pi(\theta^b)$.
  \end{enumerate}
  \item[2.] Compute $\bar \theta_{SLT} = \sum \bar w^b(\theta^b)\theta^b$, where $\bar w^b=\frac{w^b
  (\theta^b)}{\sum_{c=1}^B w^c(\theta^c)}$.
\end{enumerate}
While the SLT is a weighted average of SMD modes,
the draws of $\hat\psi^b(\theta)$ are taken from the (frequentist) asymptotic
distribution of $\hat\psi$ instead of solving the model at each $b$.  \citet{gao-hong} use a similar idea to make
draws of what we refer to as $\bar g(\theta)$ in their extension of
the BIL estimator of \citet{creel-kristensen-il} to non-separable models.

The SMD, RS, ABC, and SLT all  require  specification and simulation of the full model.
At a practical level, the  innovations $\varepsilon^1,\dots, \varepsilon^s$ used in  SMD and SLT
 are only drawn from $F_\varepsilon$ once and held  fixed across iterations.
Equivalently, the seed of the random number generator is fixed
so that the only difference in successive iterations is due to change in the parameters
to be estimated.  In contrast,
ABC  draws new innovations from $
F_\varepsilon$ each time a $\theta^{b+1}$ is
 proposed. We need to simulate $B$ sets of innovations of length $T$,
not counting those used in  draws that  are rejected, and $B$ is generally much bigger
than $S$. The SLT takes $B$ draws from an asymptotic distribution of $\hat\psi$.
 Hence even though some aspects of the  algorithms considered seem
 similar, there are subtle differences.

\section{Properties of the Estimators}
This section studies the finite sample properties of the various  estimators.
Our goal is to compare the SMD with the RS, and by implication, the infeasible
MCMC-ABC. Note that our RS is different from the original kernel based ABC methods.
To do so in a tractable way, we only consider the expansion up to order $\frac{1}{T}$.
As a point of reference, we first note that
under assumptions in
\citet{rsu-96,bao-ullah:07},
$\hat\theta_{ML}$ admits a second order expansion
\[ \hat\theta_{ML}=\theta_0+\frac{A_{ML}(\theta_0)}{\sqrt{T}}+\frac{C_{ML}(\theta_0)}{T}+o_p(\frac1T).
\]
where
$A_{ML}(\theta_0)$ is a mean-zero asymptotically normal random vector and $C_{ML}(\theta_0)$
depends on the curvature of the likelihood. These terms  are defined as
\begin{subequations}
\begin{eqnarray}
\label{eq:a-ml}
A_{ML}(\theta_0)&=& \mathbb E[\ell_{\theta\theta}(\theta_0)]^{-1} Z_{S}(\theta_0)\\
\label{eq:c-ml}
C_{ML}(\theta_0)&=& \mathbb E[-\ell_{\theta\theta}(\theta_0)]^{-1}\bigg[Z_H(\theta_0)
Z_{S}(\theta_0)-\frac{1}{2}\sum_{j=1}^ K  (-\ell_
{\theta\theta\theta_j}(\theta_0))Z_{S}(\theta_0)Z_{S,j}(\theta_0)\bigg]
\end{eqnarray}
\end{subequations}
where the normalized score $\frac{1}{\sqrt{T}} \ell_\theta(\theta_0)$
and centered Hessian $\frac{1}{\sqrt{T}}
( \ell_{\theta\theta}(\theta_0)-\mathbb E[\ell_{\theta\theta}(\theta_0)])$ converge in
distribution to the normal vectors $Z_{S}$ and $Z_H$ respectively.
The order $\frac1T$ bias  is large when
Fisher information is low.

Classical Bayesian estimators are likelihood based. Hence  the posterior mode $\hat\theta_{
BC}$ exhibits a bias similar to that of $\hat\theta_{ML}$.
However, the prior
$\pi(\theta)$ can be thought of as a constraint, or penalty since
the posterior mode maximizes  $\log p(\theta|\mathbf y)=
\log L(\theta|\mathbf y)+\log \pi(\theta)$.
Furthermore, \citet{kass-tierney-kadane}  show that
 the posterior mean deviates from the posterior mode
by  a term that depends on the second derivatives of the log-likelihood. Accordingly,
there are three sources of bias in the posterior mean $\bar\theta_{BC}$:
a likelihood component, a prior component, and a component from approximating
the mode by the mean. Hence
\begin{eqnarray*}
 \hat{\theta}_{BC} = \theta_0 + \frac{A_{ML}(\theta_0)}{\sqrt{T}} + \frac{1}{T} \bigg
 [C_{BC} (\theta_0) + \frac{\pi_\theta(\theta_0)}{\pi(\theta_0)}C^P_{BC}(\theta_0)
 + C^M_{BC}(\theta_0)\bigg] + o_p(\frac{1}{T}).
\end{eqnarray*}
Note that the  prior component is
 under the control of the researcher.

In what follows, we will show that posterior means based on auxiliary statistics
$\hat\psi$  generically have the above  representation, but the composition
of the terms differ.



\subsection{Properties of $\hat\theta_{SMD}$}

Minimum distance estimators depend on   auxiliary statistics
 $\hat\psi$. Its properties have been analyzed in \citet[Section 4.2]{newey-smith-04}
 within an empirical-likelihood framework. To facilitate
 subsequent analysis, we  follow \citet[Ch.4.4]{g-monfort-simulation} and directly
 expand $\hat \psi$ around $\psi(\theta_0)$, under the assumption that it admits a second-order expansion. In particular, since   $\hat\psi$  is $\sqrt{T}$
consistent for $\psi(\theta_0)$, $\hat\psi$ has  expansion
 \begin{equation}
\label{eq:psi-hat}
 \hat\psi=\psi(\theta_0)+\frac{\mymathcal A(\theta_0)}{\sqrt{T}}+
\frac{\mymathcal C(\theta_0)}{T}+o_p(\frac1T).
\end{equation}
It is then straightforward to show that the minimum distance estimator $\hat\theta_{MD}$ has expansion
\begin{subequations}
\begin{eqnarray}
A_{MD}(\theta_0)&=& \Big[\psi_\theta(\theta_0)\Big]^{-1} \mymathcal A(\theta_0)\label
{eq:A-MD}\\
C_{MD}(\theta_0)&=& \Big[\psi_\theta(\theta_0)\Big]^{-1} \bigg[ \mymathcal C(\theta_0)-\frac{1}
{2} \sum_{j=1}^K \psi_{\theta,\theta_j}(\theta_0)   A_{MD}(\theta_0)A_{MD,j} (\theta_0)\bigg].
\label{eq:C-MD}
\end{eqnarray}
\end{subequations}
The bias in $\hat\theta_{MD}$ depends on the curvature of the binding function
and   the bias in the auxiliary statistic $\hat\psi$,  $\mymathcal C(\theta_0)$.
Then following  \citet{grt-99}, we can analyze the SMD as follows.
In view of (\ref{eq:psi-hat}), we have,
 for each $s$:
\begin{eqnarray*}
\hat\psi^s(\theta)&=&\psi(\theta)+\frac{\mymathcal A^s(\theta)}{\sqrt{T}}+\frac{\mymathcal
C^s (\theta)}{T}+
o_p(\frac{1}{T}).
\end{eqnarray*}

The estimator   $\hat\theta_{SMD}$
satisfies $\hat\psi= \frac{1}{S}\sum_{s=1}^S \hat \psi^s(\hat\theta_
{SMD})$ and has expansion $\hat\theta_{SMD}= \theta_0+\frac{A_{SMD}(\theta_0)}{\sqrt
{T}}+\frac{C_{SMD}(\theta_0)}{T}+o_p(\frac1T)$.
Plugging it in the Edgeworth expansions gives:
\begin{eqnarray*} \psi(\theta_0)+\frac{\mymathcal A(\theta_0)}{\sqrt{T}}+\frac{\mymathcal C
(\theta_0)}
{T}+
O_p(\frac{1}{T})
=\frac{1}{S}\sum_{s=1}^S \bigg[ \psi(\hat\theta_{SMD}) +\frac{\mymathcal A^s(\hat\theta_
{SMD})} {\sqrt{T}}+\frac{\mymathcal C^s(\hat \theta_{SMD})}{T}+
o_p(\frac{1}{T}) \bigg].
\end{eqnarray*}
Expanding $\psi(\hat\theta_{SMD})$ and $\mymathcal A^s(\hat\theta_{SMD})$
around $\theta_0
$ and equating terms in the expansion of $\hat\theta_{SMD}$,
\begin{subequations}
 \begin{eqnarray}
  A_{SMD}(\theta_0)&=&\bigg[\psi_\theta(\theta_0)\bigg]^{-1}
\bigg(  \mymathcal A(\theta_0)-\frac{1}{S}\sum_{s=1}^S
\mymathcal A^s (\theta_0)\bigg) \label{eq:A-SMD}\\
 C_{SMD}(\theta_0)&=&\bigg[\psi_\theta(\theta_0)\bigg]^{-1}\bigg(\mymathcal C(\theta_0)-
\frac{1}{S}\sum_{s=1}^S \mymathcal
C^s (\theta_0)-
\big( \frac {1}{S}\sum_{s=
1}^S  \mymathcal A_\theta^s(\theta_0) \big) A_{SMD}(\theta_0) \bigg) \label
{eq:C-SMD}\\
&&-\frac{1}{2} \bigg[\psi_\theta(\theta_0)\bigg]^{-1}\sum_{j=1}^K
\psi_{\theta,\theta_j}
(\theta_0) A_{SMD}(\theta_0)A_{SMD,j}(\theta_0). \nonumber
\end{eqnarray}
\end{subequations}
The first order term can be written as $A_{SMD}=A_{MD}+\frac{1}{B}[\psi_\theta(\theta_0)]^{-1}\sum_{b=1}^B \mathbb A^b(\theta_0)$, the last term has variance of order $1/B$ which accounts for simulation noise.
Note also that $\mathbb{E}\left( \frac{1}{S}\sum_{s=1}^S \mymathcal C^s(\theta_0) \right) = \mathbb E[\mymathcal C
(\theta_0)]$. Hence, unlike the MD, $\mathbb E[C_{SMD}(\theta_0)]$ does not
depend on the bias $\mathbb{C}(\theta_0)$ in the auxiliary statistic.
In the special case when $\hat\psi$ is a consistent estimator of $\theta_0$,
 $\psi_
{\theta}(\theta_0)$ is the identity map and the term involving $\psi_{\theta\theta_j}
(\theta_0)$ drops  out. Consequently,
the SMD has no bias of order $\frac{1}{T}$ when
$S\rightarrow\infty$  and $\psi(\theta)=\theta$. In general, the bias of $\hat\theta_
{SMD}$ depends on the curvature of the binding function as
\begin{equation}
\label{eq:bicC-SMD} \mathbb E[C_{SMD}(\theta_0)]\overset{S \to \infty}{\to}-\frac{1}{2} \bigg[\psi_\theta(\theta_0)\bigg]^{-1}\sum_{j=1}^K
\psi_{\theta,\theta_j}
(\theta_0) \mathbb E\bigg[A_{MD}(\theta_0)A_{MD,j}(\theta_0)\bigg].
\end{equation}
This is an improvement over $\hat\theta_{MD}$ because as seen from (\ref{eq:C-MD}),
\begin{equation}
\label{eq:bicC-MD} \mathbb E[C_{MD}(\theta_0)]=\bigg[\psi_\theta(\theta_0)\bigg]^{-1}\mathbb{C}(\theta_0)-\frac{1}{2} \bigg[\psi_\theta(\theta_0)\bigg]^{-1}\sum_{j=1}^K
\psi_{\theta,\theta_j}
(\theta_0) \mathbb E\bigg[A_{MD}(\theta_0)A_{MD,j}(\theta_0)\bigg].
\end{equation}
The bias in $\hat\theta_{MD}$ has an additional term in $\mathbb
C(\theta_0)$.


\subsection{Properties of  $\bar\theta_{RS }$}

The convergence properties of the ABC algorithms have been well analyzed but
the theoretical properties
  of the estimates are less understood. \citet{dsjp} establish consistency of the ABC
in the case of hidden Markov models.
The analysis considers  a  scheme so that maximum likelihood estimation based on the ABC
algorithm is equivalent to exact inference under the perturbed hidden Markov scheme.
The authors find that the asymptotic bias  depends on the ABC tolerance  $\delta$.
\citet{calvet-czellar:14} provide an upper bound for the mean-squared error
of their ABC filter and study how the choice of the bandwidth affects properties
of the  filter.
Under high level conditions and adopting the  empirical likelihood
framework of \citet {newey-smith-04},
\citet{creel-kristensen-il}
show that
the infeasible BIL is second order equivalent to the MIL after bias adjustments,
while  MIL  is in turn first
order equivalent to the continuously updated GMM.
 The feasible SBIL (which is also an ABC estimator) has  additional errors compared to the BIL
 due to simulation noise and kernel smoothing, but these errors  vanish as $S\rightarrow\infty$
 for an appropriately chosen bandwidth.  \citet{gao-hong} show
that  local-regressions have better variance properties compared to kernel
estimations of the indirect likelihood. \citet{cghk:16} show that
the number of simulations can affect the parametric convergence rate and asymptotic normality
of the estimator, which is important for frequentist inference.


ABC algorithms are traditionally implemented using kernel smoothing, the first implementation
being  \citet{beaumont-zhang-balding}. The bias due to kernel smoothing
 is rigorously studied in \citet{cghk:16} under the assumption
that the draws are taken directly from the prior. Our RS is an importance sampler that
does not use kernel smoothing. Instead it uses optimization to set $\delta$
 equal to zero. This offers different insight as we   look at the bias in
 the ideal case where $\delta$ is exactly zero.

As shown above,  $\bar\theta_
{RS }$ is the weighted average of a sequence of SMD modes. Analysis of the
weights $w^b(\theta^b)$ requires an expansion of
 $\hat\psi^b_\theta
(\theta^b)$ around $\psi_\theta(\theta_0)$. From such an analysis, shown in the
Appendix, we find that
\begin{eqnarray*}
\bar
\theta_ {RS }&=&\sum_ {b=1}^B \bar w^b(\theta^b) \theta^b= \theta_0+\frac{ { A}_{RS }(\theta_0)}{\sqrt{T}}+\frac{
{C}_{RS }(\theta_0)}{T}+
o_p(\frac{1}{T})
\end{eqnarray*}
where
\begin{subequations}
\begin{eqnarray}
 A_{RS }(\theta_0)&=& \frac{1}{B}\sum_{b=1}^B A_{RS}^b(\theta_0)=\bigg[\psi_\theta(\theta_0)\bigg]^{-1}
\bigg( \mymathcal A(\theta_0) -\frac{1}{B}\sum_{b=1}^B\mymathcal A^b(\theta_0)\bigg)
\label{eq:A-rs}\\
 C_{RS }(\theta_0)&=& \frac{1}{B}\sum_{b=1}^B C_{RS}^b(\theta_0) +  \frac {\pi_\theta(\theta_0)}
{\pi(\theta_0)}\left[\frac{1}{B}\sum_
{b=1}^B  (A_{RS}^b(\theta_0)-\bar A_{RS}(\theta_0))A_{RS}^b(\theta_0) \right]
+ C^M_{RS}(\theta_0).
\label{eq:C-rs}
\end{eqnarray}
\end{subequations}

\begin{proposition}
\label{prop:prop2}
Let $\hat\psi(\theta)$ be the auxiliary statistic that admits the expansion as
in (\ref{eq:psi-hat})
and suppose that   the prior $\pi(\theta)$ is positive and continuously
differentiable around  $\theta_0$ when $dim(\hat \psi)=dim(\theta)$.
Then $ \mathbb E[A_{RS}(\theta_0)]=0$ but $\mathbb E[C_{RS }(\theta_0)]\ne 0$ for an arbitrary choice of prior.
\end{proposition}

The SMD and RS  are first
order equivalent, but
 $\bar\theta_{RS }$ has an order $\frac
{1}{T}$ bias.
The  bias, given by $C_{RS}(\theta_0)$, has three components. The
 $C^M_ {RS}(\theta_0)$ term (defined in  Appendix A)
 can be traced directly to the weights, or to the interaction of the weights with the
prior, and is a function of $A_{RS}(\theta_0)$. Some but not all the terms vanish as $B\rightarrow \infty$. The second term   will be zero if a uniform prior is chosen
since $\pi_\theta=0$. A similar result is obtained in \citet{creel-kristensen-il}.
The first term is

{\small
\begin{eqnarray*}
\frac{1}{B}\sum_{b=1}^B C_{RS}^b(\theta_0)&=&\bigg[\psi_\theta(\theta_0)\bigg]^{-1} \frac{1}{B}\sum_{b=1}^B\bigg (
\mymathcal C(\theta_0)-\mymathcal C^b(\theta_0) -\frac{1}{2}\sum_{j=1}^K \psi_
{\theta\theta_j}(\theta_0) A^b_{RS}(\theta_0)A^b_
{RS,j}(\theta_0)-\mymathcal A_{\theta}^b
(\theta_0) A^b_{RS}(\theta_0)\bigg).
\end{eqnarray*}
}
The
 term $ \mymathcal C(\theta_0)-\frac{1}{B}\sum_{b=1}^B\mymathcal C^b(\theta_0)$
 is exactly the same as in
$C_{SMD}(\theta_0)$. The  middle term involves $\psi_{\theta\theta_j}(\theta_0)$
and is zero if $\psi(\theta)=\theta$. But because the summation is over $\theta^b$ instead of $\hat \psi^s$,
\begin{equation*}\frac{1}{B}\sum_{b=1}^B \mymathcal A^b_{\theta}(\theta_0) A_
{RS}^b
(\theta_0)\overset{B\rightarrow\infty}{\rightarrow} \mathbb E[\mymathcal A_{\theta}^b(\theta_0) A_{RS}^b(\theta_0)]
\ne 0.
\end{equation*}
As a consequence
$ \mathbb E[C_{RS}(\theta_0)]\ne 0$
even when $\psi(\theta)=\theta$.
In contrast, $\mathbb E[ C_{SMD}(\theta_0)]=0$ when $\psi(\theta)=\theta$
 as seen from (\ref{eq:bicC-SMD}). The reason is that
 the
 comparable term  in $C_{SMD}(\theta_0)$ is
\begin{equation*}
\bigg(\frac{1}{S}\sum_{s=1}^S \mymathcal A_{\theta}^s (\theta_0)\bigg)
A_{SMD}
(\theta_0)\overset{S\rightarrow\infty}{\rightarrow} \mathbb  E[\mymathcal A_{\theta}^s(\theta_0)] A_{SMD}(\theta_0)=0.
\end{equation*}
 The difference
 boils down to  the fact that the SMD is the mode  of the
  average over simulated auxiliary statistics,  while
 the RS is a weighted average over the modes. As will be seen below, this difference is also present in the
 LT  and SLT and comes from averaging over $\theta^b$.
The result is based on fixing
$\delta$  at zero and holds for any $B$. Proposition \ref{prop:prop2} implies that the ideal MCMC-ABC with $\delta=0$
also has a non-negligible second-order bias.  Note that Proposition \ref{prop:prop2} is stated for the exactly identified case.
 When  $dim(\hat \psi)>dim(\theta)$, the analysis is more complicated. Essentially, when the model is overidentified, weighting is needed since all moments cannot be made equal to zero simultaneously in general. This introduces
 additional biases.  A result analogous to Proposition \ref{prop:prop2} is given in \citet{jjng-15} for the overidentified case.

In theory,  the order $\frac{1}{T}$ bias can be
removed if
$\pi(\theta)$ can be found to  put the right hand side of $C^{RS}(\theta_0)$
defined in (\ref {eq:C-rs}) to zero.
Then $\bar\theta_{RS }$ will  be second order equivalent to  SMD when $\psi
 (\theta)=\theta$ and may have a smaller bias than SMD when $\psi(\theta)\ne \theta$
 since  SMD
has a non-removable second order bias in that case. That the choice
of prior will have bias implications  for likelihood-free estimation  echoes
the findings  in the parametric likelihood setting.  \citet{ArellanoBonhomme}
show in the context of non-linear panel data models  that
the first-order bias  in Bayesian estimators can be eliminated
with a particular prior on the individual effects. \citet{bester-hansen-06} also
show that in the
estimation of parametric likelihood models,
the  order $\frac{1}{T}$ bias in the posterior mode and mean can be removed using objective Bayesian
priors.  They suggest to replace the
population quantities in a differential equation  with sample estimates.
Finding the  bias-reducing prior for the RS  involves
 solving the differential equation:
\[0= \mathbb{E}[C_{RS}^b(\theta_0)] +  \frac {\pi_\theta(\theta_0)}
{\pi(\theta_0)}\mathbb{E}[(A_{RS}^b(\theta_0)-\bar A_{RS}(\theta_0))A_{RS}^b(\theta_0)]
+\mathbb{E}[C^M_{RS}(\theta_0),\pi(\theta_0)]
\]
which has the additional dependence on $\pi$ in $C^M_{RS}(\theta_0,\pi
(\theta_0))$ that is not present in \citet{bester-hansen-06}.
A closed-form solution is available
 only for simple examples as we will see Section 6.1 below. For realistic
problems, how to find and implement the bias-reducing prior is  not a trivial problem.
A natural starting point is the  plug-in procedure of \citet{bester-hansen-06}
but little is known about its finite sample properties even in the likelihood setting for which it was developed.

This section
has studied the RS, which is the best that the
MCMC-ABC can achieve in terms of $\delta$. This enables us to make a comparison
with the SMD holding  the same $\mathsf L_2$ distance  between $\hat\psi$ and $\psi(\theta)$
at  zero by machine precision.
However, the MCMC-ABC algorithm with
$\delta>0$ will not produce draws with the same distribution as the RS. To
see the problem, suppose that the
RS draws are obtained by stopping the optimizer  before
$\|\hat\psi-\psi(\theta^b)\|$ reaches the tolerance guided by
machine precision. This is analogous to equating
 $\psi(\theta^b)$  to the pseudo estimate $\hat\psi+\delta$. Inverting the
binding function will yield an estimate of $\theta$ that depends on the random
$\delta$
in an intractable way. The RS estimate will thus have an additional bias from
$\delta\ne 0$.
By implication,
 the MCMC-ABC with $\delta>0$ will be second order equivalent to the SMD only
 after a bias adjustment even when $\psi(\theta)=\theta$.


\subsection{The Properties of LT and SLT}
The
mode of $\exp(-J(\theta))\pi(\theta)$  will inherit
the properties of a MD estimator.  However,  the quasi-posterior mean has two additional
sources of bias, one arising from the prior, and another one from   approximating
the mode by the mean.
The optimization view of $\bar\theta_{LT}$  facilitates an understanding of
these effects.
As shown in Appendix B, each draw $\theta^b_{LT}$ has expansion terms
\begin{eqnarray*}
  A_{LT}^b(\theta_0)&=&\Big[ \psi_\theta(\theta_0)\Big]^{-1} \left( \mymathcal A
  (\theta_0) - \mymathcal A_\infty^b(\theta_0)\right)\\
  C_{LT}^b(\theta_0)&=&\Big[ \psi_\theta(\theta_0)\Big]^{-1} \left( \mymathcal C
  (\theta_0)-\frac{1}{2}\sum_{j=1}^K \psi_{\theta,\theta_j}(\theta_0)(A_{LT}^b
  (\theta_0) A^b_
  {LT,j}(\theta_0)
  - \mymathcal A_{\infty,\theta}^b(\theta_0)A^b_{LT}(\theta_0)\right).
\end{eqnarray*}
Even though the LT has the same objective function as MD, simulation
noise enters both $A^b_{LT}(\theta_0)$ and $C^b_{LT}(\theta_0)$.
Compared to the
extremum estimate $\hat\theta_{MD}$,   we see that $ A_{LT}= \frac{1}{B}\sum_{b=1}^B
A_{LT}^b(\theta_0)\ne A_{MD}(\theta_0)$  and $ C_{LT}(\theta_0)\ne C_{MD}(\theta_0)$.
Although $C_{LT}(\theta_0)$ has
the same terms as $C_{RS}(\theta_0)$,
 they are  different because the LT uses the asymptotic binding function,
 and hence $A^b_ {LT}(\theta_0)\ne  A^b_{RS}(\theta_0)$.

A similar  stochastic
expansion of each $\theta^b_{SLT}$ gives:
\begin{eqnarray*}
  A_{SLT} ^b(\theta_0) &=& \Big[ \psi_\theta(\theta_0)\Big]^{-1} \left( \mymathcal
  {A}(\theta_0) - \frac{1}{S}\sum_{s=1}^S \mymathcal{A}^s(\theta_0)-\mymathcal
  {A}_\infty^b(\theta_0)\right)\\
  C_{SLT}^b(\theta_0) &=& \Big[ \psi_\theta(\theta_0) \Big]^{-1}\left( \mymathcal
  {C}(\theta_0)-\frac{1}{S}\sum_{s=1}^S \mymathcal{C}^s(\theta_0) - \frac
  {1}{2}\sum_{j=1}^K \psi_{\theta,\theta_j}(\theta_0)A_{SLT}^bA^b_{SLT,j}]\right)\\
  && -\Big[ \psi_\theta(\theta_0) \Big]^{-1}
   \left(\frac{1}{S}\sum_{s=1}^S (\mymathcal{A}_{\theta}^s(\theta_0)+\mymathcal
   {A}_{\infty,\theta}^b
  (\theta_0)) A^b_{SLT}(\theta_0)\right)
\end{eqnarray*}
Following the same argument as in the RS,
an optimally chosen prior can reduce bias,  at least in theory, but finding this prior will not be a trivial task.
 Overall, the SLT has features of the RS (bias does not depend on $\mymathcal
 {C}(\theta_0)$) and the LT (dependence on $\mymathcal{A}^b_\infty$) but is different from both.
 Because the SLT  uses simulations
to approximate
the  binding function $\psi(\theta)$,   $\mathbb E[\mathbb
C(\theta_0)-\frac{1}{S}\sum_{s=1}^S \mathbb C^s(\theta_0)]=0$. The improvement
over the LT is analogous
to the improvement of SMD over MD.  However, the $A^b_{SLT}(\theta_0)$ is
affected by
estimation of the binding function (the term with superscript $s$) and of the quasi-posterior density
(the terms with superscript $b$). This results in simulation noise with variance
of order $1/S$ plus another of order $1/B$. Note also that the SLT bias has
an additional term
\[ \frac{1}{B}\sum_{b=1}^B
 \left(\frac{1}{S}\sum_{s=1}^S (\mymathcal{A}_{\theta}^s(\theta_0)+\mymathcal
 {A}_{\infty,\theta}^b
(\theta_0)) A^b_{SLT}(\theta_0)\right) \overset{S \to \infty}{\to} \frac{1}{B}\sum_{b=1}^B
 \mymathcal
 {A}_{\infty,\theta}^b
(\theta_0) A^b_{LT}(\theta_0).\]
The main difference with the RS is that $\mymathcal{A}^b$  is replaced with
$\mymathcal{A}^b_\infty$.  For $S=\infty$ this term matches that of the LT.

\subsection{Overview}
We started this section by noting that the Bayesian posterior mean has two
components in its bias, one arising from the prior which acts like a penalty
on the objective function, and another due to approximating the mean
with the mode. We are now in a position to use the results in the foregoing
subsections to show that
 for $d$=(MD, SMD, RS, LT) and SLT and $D=$ (RS,LT,SLT) these
estimators can be represented as
\begin{equation}
\label{eq:general}
 \hat{\theta}_d = \theta_0 + \frac{A_d(\theta_0)}{\sqrt{T}} +
 \frac{C_d (\theta_0)}{T} +\frac{\mathbbm{1}_{d\in D}}{T}
\bigg[ \frac{\pi_\theta(\theta_0)}{\pi(\theta_0)}C^P_d(\theta_0))
 + C^M_d(\theta_0)\bigg] + o_p(\frac{1}{T})
 \end{equation}
 where with $      A_d^b(\theta_0) = [\psi_\theta(\theta_0)]^{-1} \Big( \mymathcal A
      (\theta_0) - \mymathcal A_d^b(\theta_0) \Big)$,
 \begin{eqnarray*}
  A_d(\theta_0) &=& [\psi_\theta(\theta_0)]^{-1} \Big( \mymathcal A
      (\theta_0) -\frac{1}{B}\sum_{b=1}^B \mymathcal A_d^b(\theta_0) \Big)\\
      C_d(\theta_0) &=& [ \psi_\theta(\theta_0) ]^{-1} \Big( \mymathcal{C}(\theta_0)-\mymathcal{C}_d(\theta_0) - \frac{1}{2}\sum_{j=1}^K \psi_{\theta,\theta_j}(\theta_0) A^b_d(\theta_0)A^b_{d,j}(\theta_0) - \mymathcal{A}^b_{d,\theta}A^b_d(\theta_0)\Big)\\
      C^P_d(\theta_0)&=& \frac{1}{B}\sum_{b=1}^B (A^b_d(\theta_0) - A_d(\theta_0))A^b_d
      (\theta_0),
\end{eqnarray*}
The term $C^P_d(\theta_0)$ is a bias directly due to the prior.
The term $C^M_d(\theta_0)$, defined in the Appendix, depends on $A_d(\theta_0)$,
the curvature of the binding function, and their interaction with the prior.
Hence at a general level,
the estimators can be distinguished by whether or not Bayesian computation tools
are used, as the indicator function is null only for the two frequentist estimators
(MD and SMD). More fundamentally,
the estimators differ because of
  $A_d (\theta_0)$ and $C_d(\theta_0)$, which in turn depend on
  $\mathbb A^b_d(\theta_0)$  and $\mathbb C_d(\theta_0)$.
We compactly summarize the differences  as follows:
\begin{center}
  \begin{tabular}{c|ccccc} \hline \hline
     $d$ & $\mymathcal{A}^b_d(\theta_0)$ &  $\mymathcal C_d(\theta_0)$ & $\text
     {var}(\mymathcal{A}_d(\theta_0))$ & $\mathbb{E}[\mymathcal C(\theta_0)-\mymathcal
     C_d(\theta_0)]$\\ \hline
    MD  & 0 & 0 & 0 & $\mathbb{E}[\mymathcal C(\theta_0)]$\\
    LT  & $\mymathcal{A}_\infty^b(\theta_0)$ & 0 & $\frac{1}{B}\var[\mymathcal A^b_\infty(\theta_0)]$
    & $\mathbb{E}[\mymathcal C(\theta_0)]$\\
    RS & $\mymathcal{A}^b(\theta_0)$ & $\frac{1}{B}\sum_{b=1}^B \mymathcal{C}^b(\theta_0)$
    & $\frac{1}{B}\text{var}[\mymathcal{A}^b(\theta_0)]$ & $0$\\
    SMD & $\frac{1}{S}\sum_{s=1}^S \mymathcal{A}^s(\theta_0)$ & $\frac{1}
    {S}\sum_{s=1}^S \mymathcal{C}^s(\theta_0)$ & $\frac{1}{S}\text{var}[\mymathcal
    {A}^s(\theta_0)]$ & $0$\\
    SLT & $\mymathcal A_{SMD}(\theta_0)+\mymathcal A^b_{LT}(\theta_0)$ & $\frac{1}{S}\sum_
    {s=1}^S \mymathcal{C}^s(\theta_0)$ & $\text{var}[\mymathcal{A}_\text{SMD}(\theta_0)]+\text
    {var}[\mymathcal{A}_\text{LT}(\theta_0)]$ & $0$\\
    \hline \hline
  \end{tabular}
\end{center}

The MD is the only estimator that is optimization based and does not involve
simulations. Hence it does not depend on $b$ or $s$ and has no simulation
noise. The SMD does not
depend on $b$ because the optimization problem is solved only once.
The LT  simulates from the asymptotic binding function. Hence
its errors are associated with parameters of the asymptotic distribution.

The MD and LT have a bias due to asymptotic approximation of the binding function.
In such cases,  \cite{cabrera-fernholz} suggest to adjust an initial estimate $\tilde\theta$
 such
  that if the new estimate $\hat\theta$ were the true value of $\theta$, the
  mean of the original estimator equals the observed value $\tilde\theta$.
  Their {\em target estimator} is the $\theta$ such that
  $\mathbb E_{\mathcal P_{\theta}}[\hat \theta]=\tilde \theta$.
While the bootstrap directly estimates the
  bias, a target estimator corrects for the bias implicitly.     \citet{cabrera-hu} show that  the
  bootstrap estimator corresponds to the first step of a target estimator.
  The latter improves upon the bootstrap estimator by providing more iterations.

An   auxiliary statistic based
  target estimator  is  the $\theta$ that  solves $\mathbb E_{\mathcal P
  _\theta}[
\hat\psi (\mathbf y( \theta))] =\hat\psi(\mathbf y(\theta_0))$.
It  replaces the asymptotic binding function $\lim_{T\rightarrow\infty}
\mathbb E[\hat\psi(\mathbf y(\theta_0))]$ by
$\mathbb E_{\mathcal P
  _\theta}[ \hat\psi (\mathbf y( \theta))]$ and approximates the expectation
  under $\mathcal P_\theta$
 by stochastic
expansions. The SMD and SLT can be seen as target estimators that
approximate    the   expectation by    simulations.
Thus, they  improve upon the MD estimator even when the binding function is tractable
and is especially appealing when it is not. However,  the improvement
in the SLT is partially offset by having to approximate the mode by the
mean.

\section{Two Examples} \label{sec:Example}

The preceding section can be summarized as follows. A posterior mean computed through auxiliary statistics
generically has a component due to the prior, and a component due to the approximation
of the mode by the mean. The binding function is better approximated by simulations
than asymptotic analysis. It is possible for simulation
estimation to perform better than $\hat\psi_{MD}$ even if $\psi(\theta)$  were analytically
and computationally tractable.

In this section, we first illustrate the  above findings
 using  a simple analytical example.
 We then evaluate the properties of the estimators using the
dynamic panel model with fixed effects.

\subsection{An Analytical Example}
We consider the simple DGP   $y_i\sim N(
 m,\sigma^2)$.
 The parameters of the model are $\theta=(m,\sigma^2)^\prime$.
We  focus on $\sigma^2$ since the estimators have more interesting properties.

The MLE of $\theta$ is
\[\hat m=\frac{1}{T}\sum_{t=1}^T y_t, \quad\quad \hat\sigma^2= \frac{1}{T}\sum_{t=1}^T (y_t-\bar y)^2.\]

While the posterior distribution is dominated by the likelihood
in large samples,   the effect
of the prior is not negligible in small samples. We therefore begin with a analysis
of the effect of the prior on the posterior mean and mode in Bayesian analysis. Details of the calculations are provided in Appendix D.1.

We consider the  prior $\pi(m,\sigma^2)= (\sigma^2)^{-\alpha} \mathbbm I_{\sigma^2>0} $, $\alpha>0$ so that
the log posterior distribution is
\[ \log p(\theta|y)=\log p(\theta|\hat m,\hat\sigma^2 )\propto \frac{-T}{2}\bigg
[\log (2\pi \sigma^2) -\alpha \log \sigma^2
- \frac{1}{2\sigma^2}\sum_{t=1}^T (y_t-m)^2\bigg]\mathbbm I_{\sigma^2>0}.\]
The  posterior mode and mean of $\sigma^2$ are
$  \sigma^2_{mode}=   \frac{T \hat\sigma^2}{T+2\alpha} $
and $  \sigma^2_{mean} = \frac{T\hat\sigma^2}{T+2\alpha-5}.$
respectively. Using the fact that $E[\hat\sigma^2]=
\frac{(T-1)}{T}\sigma^2$,
we can evaluate $\sigma^2_{mode}$, $\sigma^2_{mean}$  and their expected
 values for different $\alpha$.
 \begin{table}[ht]
 \caption{Mean $\bar\theta_{BC}$ vs. Mode $\hat\theta_{BC}$}
 \label{tbl:mode}
\begin{center}
\begin{tabular}{c|cc|ccc} \hline \hline
$\alpha$ & $\bar\theta_{BC}$ & $\hat\theta_{BC}$ & $\mathbb{E}[\bar\theta_{BC}]$ &
$\mathbb{E}[\hat\theta_{BC}]$ \\ \hline \hline
0 & $\hat\sigma^2\frac{T }{T-5} $ & $\hat\sigma^2$ & $\sigma^2\frac{T-1}{T-5}$
& $\sigma^2\frac{T-1}{T}$ \\
1 & $\hat\sigma^2\frac{T }{T-3} $ & $\hat\sigma^2\frac{T}{T+2}$ & $\sigma^2\frac
{T-1}{T-3}$ & $\sigma^2\frac{T-1}{T+2}$ \\
2 & $\hat\sigma^2\frac{T}{T-1} $ & $\hat\sigma^2\frac{T}{T+4}$
 & $\sigma^2$ & $\sigma^2\frac{T-1}{T+4}$ \\
3 & $\hat\sigma^2\frac{T }{T+1} $ & $\hat\sigma^2\frac{T}{T+6}$ & $\sigma^2\frac
{T-1}{T+1}$
& $\sigma^2\frac{T-1}{T+6}$\\
\hline \hline
\end{tabular}
\end{center}
\end{table}
Two features are of note. For a given prior (here indexed by $\alpha$), the mean does not coincide
with the mode.
Second, the statistic (be it mean or mode) varies with $\alpha$. The
Jeffrey's prior corresponds to $\alpha=1$, but the bias-reducing prior is $\alpha=2$.
In the Appendix, we  show
that the bias reducing prior for this model is $\pi^R(\theta)\propto
\frac{1}{\sigma^4}$.

Next, we consider  estimators based on auxiliary statistics:
\[ \hat\psi(\mathbf y)^\prime
 =\begin{pmatrix} \hat m &  \hat\sigma^2
 \end{pmatrix}.
\]
As these are  sufficient statistics, we can also consider (exact)
likelihood-based Bayesian inference.
For SMD estimation, we let
 $(\hat m_S, \hat \sigma^2_S)=(\frac{1}{S}\sum_{s=1}^S \hat m^
s, \frac{1}{S}\sum_
{s=1}^S \hat \sigma^{2,s})$.
The LT quasi-likelihood using the variance of preliminary estimates of $m$ and $\sigma
^2$ as weights is:
\[\exp(- J(m,\sigma^2)) = \exp\bigg(-\frac{T}{2} \bigg[ \frac{(\hat m-m)^2}{\hat\sigma^2} + \frac{
(\hat\sigma^2-\sigma^2)^2}{2\hat\sigma^4}\bigg]\bigg).
\]
The LT posterior distribution is $p(m,\sigma^2|\hat m,\hat
\sigma^2)\propto \pi(m,\sigma^2)\exp(-J(m,\sigma^2))$. Integrating out $m$ gives $p(\sigma^2|\hat
m,\hat\sigma
^2 )$.
We consider
a flat prior $\pi^U(\theta) \propto \mathbbm I_{\sigma^2 \geq 0}$ and   the bias-reducing prior $\pi^R(\theta) \propto 1/\sigma^4\mathbbm I_{\sigma^2 \geq 0}$.
The RS  is the same as the SMD under a bias-reducing prior. Thus,
\begin{eqnarray*}
\hat\sigma^2_{SMD} &=& \frac{\hat\sigma^2}{\frac{1}{ST}\sum_{s=1}^
  S \sum_{t=1}^T (e_t^s- \bar e^s)^2}
  \\    \hat\sigma^{2,R}_{RS} &=&  \frac{\hat
  \sigma^2}{\frac{1}{BT}\sum_{b=1}^B\sum_{t=1}^T (e_t^b- \bar e^b)^2}\\
  \hat\sigma^{2,U}_{RS} &=&\sum_{b=1}^B \frac{\frac{\hat{\sigma}^2}{[\sum_
  {t=1}^T(e_t^b-\bar{e}^b)^2/T]^2}}{\sum_{b^\prime=1}^B \frac{1}{\sum_{t =1}^T
  (e_t^{b^\prime}-\bar{e}^{b^\prime})^2/T}}.
\end{eqnarray*}
For completeness,  the parametric Bootstrap  bias corrected estimator $\hat\sigma^2_{\text{Bootstrap}}=2\hat\sigma^2
 - \mathbb{E}_{\text{Bootstrap}}
 (\hat\sigma^2)$ is also considered:
\begin{eqnarray*}
  \hat\sigma^2_{\text{Bootstrap}} & =& 2\hat\sigma^2 - \hat\sigma^2 \frac{T-1}{T} = \hat\sigma^2(1+\frac{1}{T}).
\end{eqnarray*}
$\mathbb{E}_{\text{Bootstrap}}(\hat\sigma^2)$ computes the expected value of
the estimator replacing the true value $\sigma^2$ with $\hat \sigma^2$, the
plug-in estimate.
In this example the bias can be computed analytically since $\mathbb{E}(\hat \sigma^2(1+\frac{1}{T}))=\sigma^2(1-\frac{1}{T})(1+\frac{1}{T})=\sigma^2(1-\frac{1}{T^2}).$
While the bootstrap does not involve inverting the binding function,
  this computational  simplicity comes at the cost of
adding a higher order bias term (in $1/T^2$).

\begin{figure}[ht]
\caption{ABC vs. RS  Posterior Density}
\label{fig:fig1}
\begin{center}
\includegraphics[width=6in,height=3.5in]{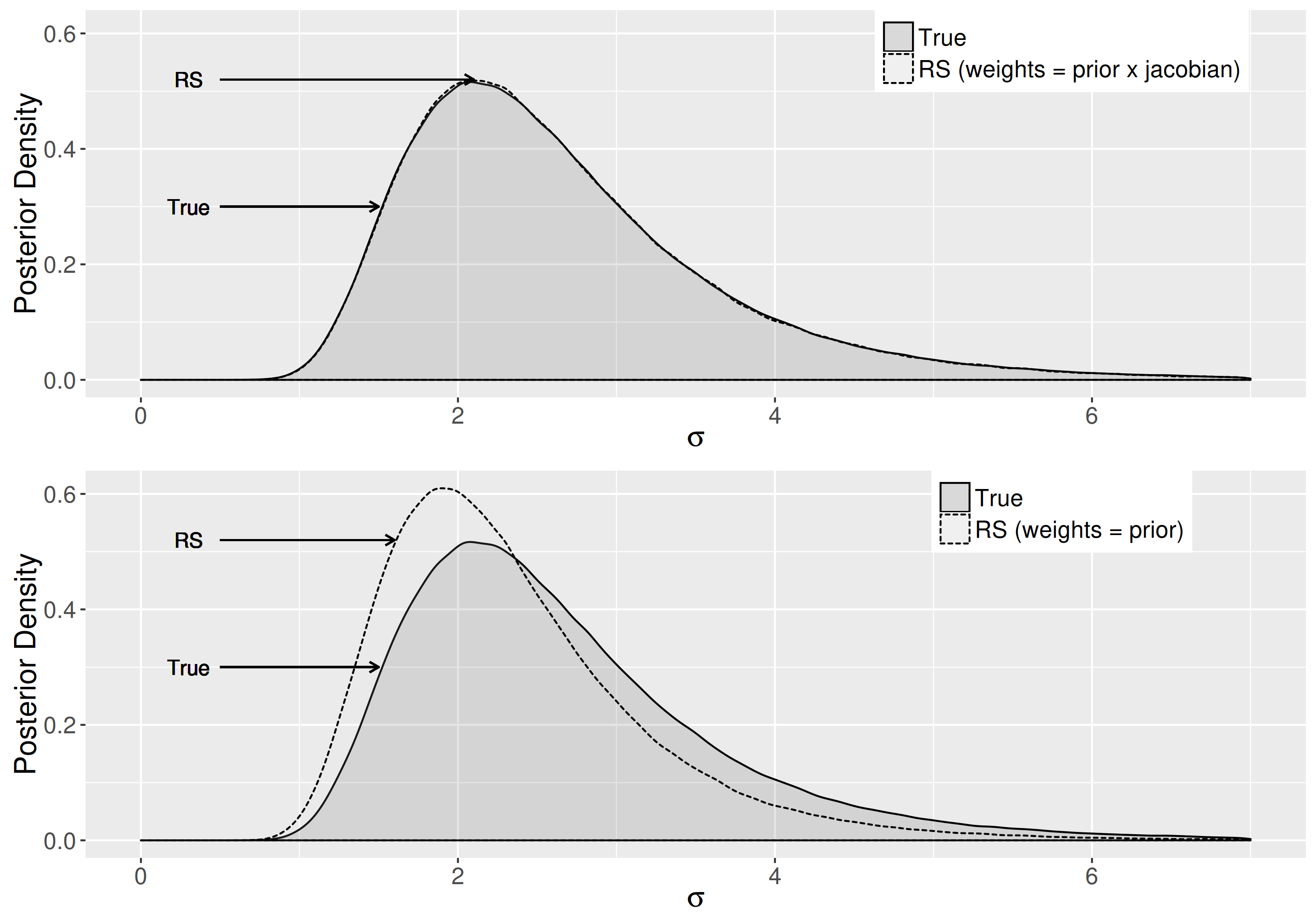}
\end{center}
\end{figure}

 A main finding of this paper is that the reverse sampler  can replicate draws from
 $p^*_{ABC}(\theta_0)$,  which in turn equals the Bayesian posterior distribution if
$\hat\psi$ are sufficient statistics.
The weight for each SMD estimate is the prior
times  the Jacobian. To illustrate the importance of the Jacobian
transformation, the top panel of Figure \ref{fig:fig1} plots the Bayesian/ABC posterior
distribution and the one obtained from the reverse sampler. They are
indistinguishable. The bottom panel shows  an incorrectly constructed reverse
sampler that does not apply the Jacobian transformation.  Notably, the two
distributions are not the same.

\begin{table}[ht]
\caption{Properties of the Estimators}
\label{tbl:simple}
\begin{center}
 \begin{tabular}{ll|l|l|ll} \hline \hline
 Estimator & Prior & $\mathbf E[\hat\theta]$  & Bias & Variance \\ \hline \hline
 $\hat\theta_{ML}$ &-&
   $\sigma^2 \frac  {T-1}{T}$  & $-\frac{\sigma^2}{T}$ & $2\sigma^4\frac{T-1}{T
   ^2} $
  \\
  $\bar\theta_{BC}$ & 1 & $\sigma^2\frac{T-1}{T-5} $ & $\frac{2\sigma^2}{T-5} $ & $2\sigma^4\frac
  {T-1}{(T-5)^2}$
  \\
  $\bar\theta^R_{BC}$ & $1/\sigma^4$ & $\sigma^2$ & 0  & $2\sigma^4 \frac{1}{T-1}$
  \\
$\bar\theta^U_{RS}$ & 1 & $\sigma^2\frac{T-1}{T-5} $ & $\frac{2\sigma^2}{T-5} $ & $2\sigma^4\frac
  {T-1}{(T-5)^2}$
  \\
  $\bar\theta^R_{RS }$ & $\frac{1}{\sigma^4}$ & $\sigma^2\frac{B(T-1)}{B(T-1)-2}$ & $\frac{2\sigma^2}{B(T-1)-2}$ &
  $  2\sigma^4 \frac{\kappa_1}{T-1}$
  \\
  $\hat\theta_{SMD}$ & - &  $\sigma^2\frac{S(T-1)}{S(T-1)-2}$ & $\frac{2\sigma^2}{S(T-1)-2}$ &
$  2\sigma^4 \frac{\kappa_1}{T-1}$
 \\
  $\bar\theta^U_{LT}$ & 1 & $\sigma^2\frac{T-1}{T}(1+\kappa_{LT})$ & $\sigma^2
  \frac  {T-1}{T}\kappa_{LT}-\frac{\sigma^2}{T}$ &
  $2\sigma  ^4 \frac{T-1}{T^2}(1+\kappa_{LT})^2$&
  \\ $\hat\theta^U_{SLT}$ & 1 & $\sigma^2\frac{S(T-1)}{S(T-1)-2}+\kappa_{SLT}$
  & $ \frac{\sigma^2}{S(T-1)-2}$+$\sigma^2\frac{T-1}{T}\mathbb E[\kappa_{SLT}]$ &
  $2\sigma^4\frac{\kappa_
  {LT}
  }{T-1}+\Delta_{SLT}$\\

  $\hat\theta_{\text{Bootstrap}}$ & - & $\sigma^2(1-\frac{1}{T^2})$ &  $\frac{-\sigma^2}{T^2}$ &  $2\sigma^4 \frac{T-1}{T^2}(1+\frac{1}{T})^2$ &
 \\
  \hline \hline
  \end{tabular}

\end{center}
\footnotesize{Notes to Table 2: Let $M(x)=\frac{\phi(x)}{1-\Phi(x)}$ be the Mills ratio.
\begin{itemize}
\item[i] $\kappa_1(S,T)= \frac{(S(T-1))^2(T-1+S(T-1)-2)}{(S(T-1)-2)^2(S(T-1)-4)}>
1$, $\kappa_1$ tends to one as $B,S$  tend to infinity.
\item[ii] $\kappa_{LT}=c_{LT}^{-1}M(-c_{LT}), $ $c_{LT}^2=\frac{T}{2}$, $\kappa_
{LT}\rightarrow 0$ as $T\rightarrow\infty$.
\item[iii] $\kappa_{SLT}=\kappa_{LT}\cdot S\cdot T \cdot \text{Inv}\chi^2_{S(T-1)}$,
 $\Delta_{,SLT}= 2\sigma^4\var  (\kappa_
  {SLT})+4\sigma^4\frac{T-1}{T^2}\cov(\kappa_{SLT},S\cdot T \text{Inv}\chi^2_{S(T-1))})$.
\end{itemize}
 }
\end{table}


 The properties of the estimators are summarized in Table \ref{tbl:simple}.
 It should be  reminded that increasing
$S$ improves the approximation of the binding function in SMD estimation
while increasing $B$ improves the approximation to the target distribution
in Bayesian type estimation.
For fixed $T$, only the Bayesian estimator with the bias reducing prior
is unbiased.
 The SMD and RS (with bias reducing prior)
have the same bias and mean-squared error
 in agreement with the analysis in the previous section.
These two estimators  have smaller
errors than the RS estimator with a uniform prior.
The SLT  posterior mean  differs from
that of the SMD by $\kappa_ {SLT}$ that is  not mean-zero.
This term, which is a function of the
Mills-ratio,  arises as a consequence of the fact that the $\sigma^2$ in SLT are drawn
from the normal distribution and then truncated to ensure positivity.

\subsection{The Dynamic Panel Model with Fixed Effects}
The dynamic panel model $y_{it}= \alpha_i + \rho y_{it-1} +  \sigma e_{it}$ is known
to be severely biased when $T$ is small because the unobserved heterogeneity
$\alpha_i$ is imprecisely estimated. Various approaches  have been
suggested to improve the precision of the least squares dummy variable (LSDV) estimator
$\hat\beta$.\footnote{See  \citet{hsiao-book}
for  a detailed account  of this incidental parameter problem.} An interesting approach, due to
\citet{gpy}, is to exploit the bias reduction properties of the indirect inference
estimator. Using the dynamic panel model as auxiliary equation, i.e. $\psi
(\theta)=\theta$, the authors reported
estimates of $\beta$  that are sharply more accurate than the LSDV, even when an exogenous regressor and a linear
trend is added to the model. Their simulation experiments
 hold $\sigma^2$
fixed.  We reconsider their exercise but also estimate $\sigma^2$.

With $\theta=(\rho,\beta,\sigma^2)^\prime$,
we simulate data from the model:
\begin{align*}
y_{it}=\alpha_i + \rho y_{it-1} + \beta x_{it} + \sigma\varepsilon_{it}.
\end{align*}
Let $A=I_T-1_T1_T^{\prime}/T$
$\underline{A}=A \otimes I_T$,
 $\underline{y}=\underline A \; vec
(y), \underline{y}_{-1}=\underline A\; vec(y_{-1}), \underline{x}=\underline A \;vec
(x)$, where $y_{-1}$ are the lagged $y$. For this model, Bayesian inference is possible since  the likelihood in de-meaned data is
 \[ L( \underline {\mathbf y},\underline {\mathbf  x}|\theta)=  \frac{1}{\sqrt{2\pi|\sigma^2\Omega|}^N}\exp \left( -\frac{1}{2\sigma^2} \sum_{i=2}^N (\underline y_i - \rho \underline y_{i,-1} - \beta \underline x_i)^\prime \Omega^{-1} (\underline y_i - \rho \underline y_{i,-1} - \beta \underline x_i)\right)\]
 where $\Omega = I_{T-1}-1_{T-1}1_{T-1}^{\prime}/T $.
We  use the following moment conditions for MD estimation:
\begin{align*}
\bar g(\rho,\beta,\sigma^2)=
\begin{pmatrix} \underline{y}_{-1}(
 \underline {y}- \rho \underline{y}_{-1} -\beta \underline{x})\\
\underline{x} (\underline {y}- \rho \underline{y}_{-1} -\beta \underline{x})\\
(\underline {y}- \rho \underline{y}_{-1} -\beta \underline{x})^2 - \sigma^2(1-1/T)
\end{pmatrix}.
\end{align*}
with $\bar g(\hat\rho,\hat\beta,\hat\sigma^2)=0$. The simulated quantity  $\bar g_S(\theta)$ for
SMD
and $\bar g^b(\theta)$ for ABC are defined analogously.
The MD estimator in this case is also the LSDV. The auxiliary estimates for
 the ABC, RS, SLT and SMD  are  the LSDV estimates.
 Recall that while
the weighting matrix $W$ is irrelevant to finding the mode in exactly identified
models, $W$ affects computation of the posterior mean.  We use $W = (\frac{1}{NT} \sum_
{i,t} g_{it}^\prime g_{it} -  \bar{g}^\prime \bar{g})^{-1}$ for LT, MCMC-ABC, and SMD.
 The prior is
$\pi(\theta)=
\mathbbm I_{\sigma^2 \geq 0, \rho \in [-1,1], \beta \in \mathbb{R}}$.  Since  the demeaned data are used in LSDV estimation, the estimates are invariant to the specification of the fixed effects. Accordingly,
 we set them to zero both in the assumed DGP and the auxiliary model.   The
innovations $\varepsilon^s$ used to simulate the auxiliary model and to construct $\hat\psi^s$ are drawn
from the standard normal distribution once and held fixed.

\begin{center}
\begin{figure}[th]
\centering
\caption{Frequentist, Bayesian, and Approximate Bayesian Inference
for $\rho$} \label{fig:Posteriors1}

 \includegraphics[width=6.0in,height=2.55in]{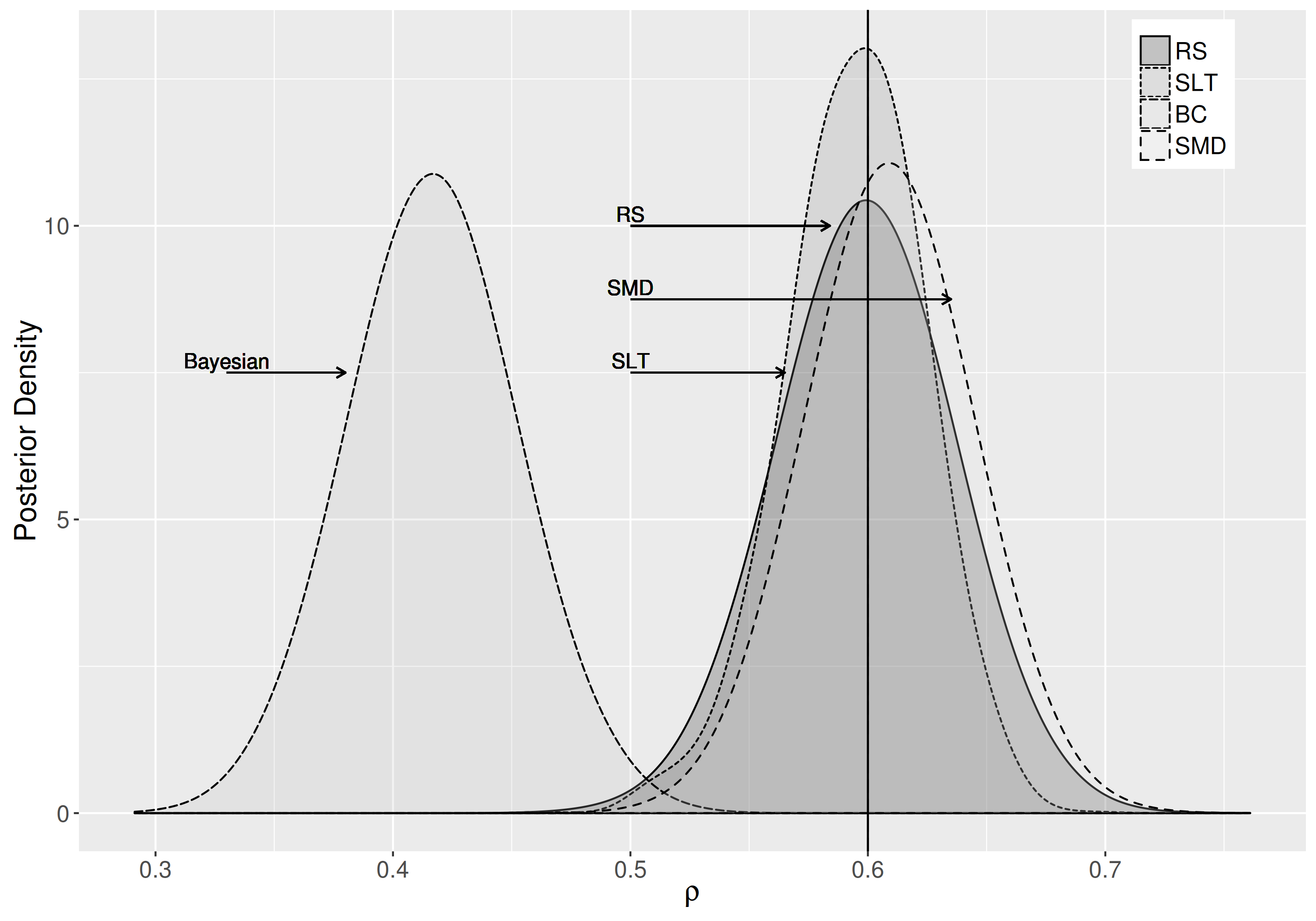}

{\footnotesize $p_{BC}(\rho|\hat\psi)$ is the likelihood based Bayesian posterior
distribution,

$p_{SLT}(\rho|\hat\psi)$ is the Simulated Laplace type quasi-posterior
distribution.

$p_{RS}(\rho|\hat\psi)$ is the approximate posterior distribution
based on the RS .

The frequentist distribution of $\hat\theta_{SMD}$ is estimated by $\mathcal
N(\hat\theta_{SMD},\hat{\var}(\hat\theta_{SMD}))$.}

\end{figure}
\end{center}

\begin{table}[ht]
  \caption{Dynamic Panel $\rho=0.6,\beta=1,\sigma^2=2$}
\label{tbl:table2}
\begin{center}
Mean over 1000 replications

\begin{tabular}{rrrrrrrrrr}
 \hline \hline
&&  MLE  & LT & SLT & SMD & $\frac{\text{MCMC}}{\text{ABC}}$ &  RS & Boot \\
 \hline \hline
 & Mean & 0.419 & 0.419 & 0.593 & 0.598 & 0.544 & 0.599 & 0.419 \\
   $\hat \rho:$ &SD & 0.037 & 0.037 & 0.038 & 0.035 & 0.036 & 0.035 & 0.074 \\
   &Bias &  -0.181 & -0.181 & -0.007 & -0.002 & -0.056 & -0.001 & -0.181 \\     \hline \hline
& Mean & 0.940 & 0.940 & 0.997 & 1.000 & 0.974 & 1.000 & 0.940 \\
     $\hat \beta:$&SD & 0.070 & 0.071 & 0.073 & 0.073 & 0.075 & 0.073 & 0.139 \\
   &Bias & -0.060 & -0.060 & -0.003 & 0.000 & -0.026 & 0.000 & -0.060 \\   \hline \hline
    &  Mean & 1.869 & 1.878 & 1.973 & 1.989 & 1.921 & 2.099 & 1.869 \\
  $\hat \sigma^2:$ &SD & 0.133 & 0.146 & 0.144 & 0.144 & 0.149 & 0.152 & 0.267  \\
   &Bias & -0.131 & -0.122 & -0.027 & -0.011 & -0.079 & 0.099 & -0.131 \\   \hline \hline
   & S & -- & -- & 500 & 500 & 1 & 1 & --\\
   & B & -- & 500& 500 & -- & 500 & 500 & 500\\ \hline \hline
\end{tabular}\\ \end{center}
Note: MLE=MD. The  MCMC-ABC uses $\delta_\text{ABC}=0.10$.

\end{table}

Table \ref{tbl:table2} reports results from 5000 replications for $T=6$ time
periods and $N=100$ cross-section units, as in
\citet {gpy}.  Both $\hat\rho$ and
$\hat\sigma^2$ are significantly biased.
The LT is the same
as the MD except that it is computed
 using Bayesian tools. Hence its  properties are similar to the MD. The simulation
 estimators have much improved properties.
The properties of $\bar\theta_{RS }$ are similar to those
 of the SMD. Figure \ref{fig:Posteriors1} 
 illustrates for one simulated dataset how the posteriors for RS /SLT are shifted towards the true value
 compared to the one based on the direct likelihood.

\begin{figure}[ht]
  \centering
  \caption{MCMC-ABC vs. RS  Posterior Density} \label{fig:bandwidth}
  \includegraphics[scale=0.5]{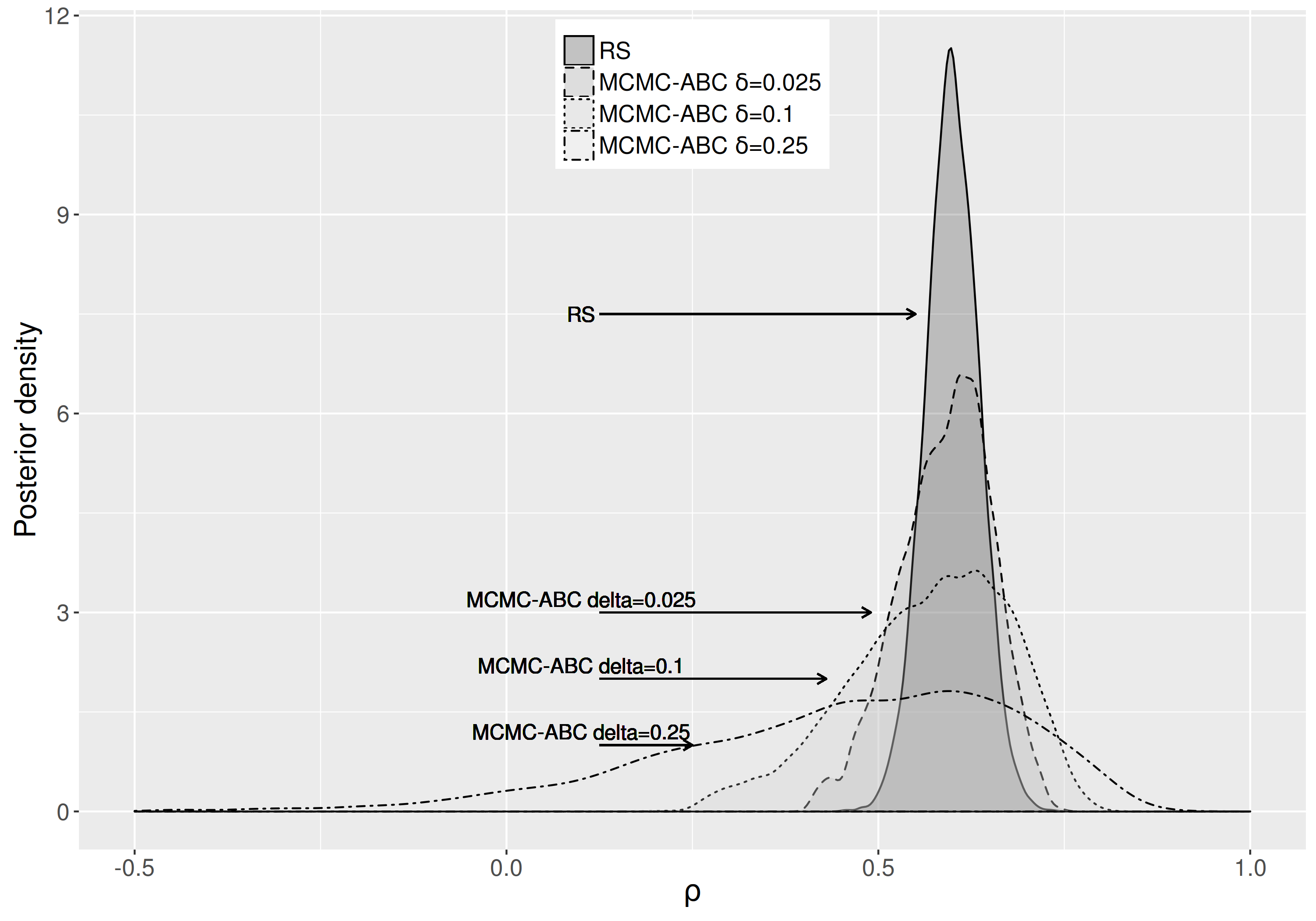}
  \end{figure}

The MCMC-ABC results in Table \ref{tbl:table2} are for $\delta=0.10$ which
has an acceptance rate of 0.58. These estimates are clearly
more precise than  MLE  but more biased than SMD or RS.
  The dependence of MCMC-ABC on $\delta$ is investigated in further detail in \citet{jjng-15}. In brief,  when we set
 $\delta=0.25$, we achieve an acceptance ratio of 0.72 but the estimates are severely biased,
  as shown in Figure  \ref{fig:bandwidth}.  Bias similar to SMD and RS can be obtained
  if we set $\delta$ to 0.025. But the corresponding acceptance rate is 0.28, meaning that
the MCMC-ABC needs at least three times more draws than the RS for a comparable level of bias.
 The choice of $\delta$ is more important for the properties of MCMC-ABC than  the RS which
 associates   $\delta$ with the tolerance of optimization.

\clearpage



\section{Conclusion}

Different disciplines have developed different
estimators to overcome  the limitations posed
by an intractable likelihood. These estimators share many similarities: they
rely on auxiliary statistics and  use simulations to approximate quantities
that have no closed form expression.  We
suggest an optimization   framework that helps understand
the estimators from the perspective of classical minimum distance estimation.
All
estimators are first-order equivalent as
$S\rightarrow\infty$ and $T\rightarrow\infty$ for any choice of $\pi(\theta)$.
Nonetheless, up to order $1/T$, the estimators are distinguished by biases
due to the prior and approximation
of the mode by the mean, the very two features that distinguish
Bayesian and frequentist estimation.



We have only considered  regular problems when $\theta_0$ is in the interior
of $\Theta$ and the objective function is differentiable. When these conditions
fail, the  posterior is no longer
asymptotically normal around the MLE with variance equal to the inverse of the
Fisher Information Matrix.
Understanding the
properties of these estimators under non-standard conditions is the subject for
future research.

\newpage
\section*{Appendix}
\setcounter{equation}{0}
\renewcommand{\theequation}{A.\arabic{equation}}
\small
The terms $\mymathcal A(\theta)$ and $\mymathcal C(\theta)$ in $\hat\theta_
{MD}$ are derived
for the just identified case as follows.
Recall that $\hat{\psi}$ has a second order expansion:
\begin{equation}
\hat{\psi} = \psi(\theta_0) + \frac{\mymathcal {A}(\theta_0)}{\sqrt{T}} + \frac{\mymathcal {C}(\theta_0)}{T}+ o_p(\frac{1}{T}).
\label{eq:A1}
\end{equation}
Now $\hat{\theta} = \theta_0 + \frac{A(\theta_0)}{\sqrt{T}}+
\frac{C(\theta_0)}{T} + o_p(\frac{1}{T}).
$ Thus expanding  $\psi(\hat\theta)$ around $\hat\theta=\theta_0$:
\begin{align*}
\psi(\hat\theta)&=\psi \left( \theta_0 + \frac{A(\theta_0)}{\sqrt{T}}+ \frac
{C(\theta_0)}{T} + o_p(\frac1T) \right) \\
&=\psi(\theta_0) +  \psi_\theta(\theta_0)\left( \frac{A(\theta_0)}{\sqrt{T}}+ \frac{C(\theta_0)}{T} + o_p(\frac1T)  \right) + \frac{1}{2T} \sum_{j=1}^K
\psi_{\theta,\theta_j} (\theta_0) A(\theta_0)A_j(\theta_0) + o_p(\frac1T).
\end{align*}
Equating with $
\psi(\theta_0) + \frac{\mymathcal {A}(\theta_0)}{\sqrt{T}} +
\frac{\mymathcal {C}(\theta_0)}{T} + o_p(\frac1T)
$ and solving for $A,C$ we get:
\begin{align*}
A(\theta_0) &= \Big[  \psi_\theta(\theta_0) \Big]^{-1} \mymathcal {A}(\theta_0) \\
C(\theta_0) &= \Big[  \psi_\theta(\theta_0) \Big]^{-1} \left( \mymathcal
{C}(\theta_0) - \frac{1}{2}\sum_{j=1}^K
\psi_{\theta,\theta_j} (\theta_0) A(\theta_0)A_j(\theta_0) \right).
\end{align*}
For estimator specific  $ A_d^b$ and $ a_d^b$, define $a^b_d = \text{trace}([\psi_\theta(\theta_0)]^{-1}[\sum_{j=1}^K \psi_{\theta,\theta_j}(\theta_0)A^b_{d,j}(\theta_0)+\mymathcal{A}^b_{d,\theta}(\theta_0)])$,
{\small
\begin{eqnarray}
C^M_d(\theta_0)&=&
2\frac{\pi_\theta(\theta_0)}{\pi(\theta_0)}\bar A_d(\theta_0)\bar a_d
(\theta_0)\theta_0 - \bar a_d (\theta_0)^2\theta_0
- \left[ \frac{\pi_\theta(\theta_0)\pi_\theta(\theta_0)^\prime}{\pi(\theta_0)^2} \right]
\bar A_d(\theta_0)^\prime \bar A_d(\theta_0)\theta_0\nonumber \\ &&-
 \frac{1}{B}\sum_{b=1}^B
(a^b_d(\theta_0)-\bar a_d(\theta_0))A^b_d(\theta_0).
\label{eq:vbar}
\end{eqnarray}
}
Where $\bar a_d = \frac{1}{B}\sum_{b=1}^B a^b_d$, $\bar A_d$ is defined analogously.
Note that $\bar a(\theta_0)\rightarrow 0$ as $B\rightarrow\infty$ if $\psi(\theta)=\theta$ and the
first two terms drop out.


\setcounter{section}{0}
\renewcommand{\thesection}{A.\arabic{section}}
\section{Proof of Proposition \ref{prop:prop2}, RS}
To prove Proposition \ref{prop:prop2}, we need an expansion for $\hat\psi^b(\theta^b)$ and
the weights using
\begin{eqnarray}
  \theta^b &= \theta_0 + \frac{A^b(\theta_0)}{\sqrt{T}}+\frac{C^b(\theta_0)}{T}+o_p(\frac{1}{T}).
   \label{eq:A2}
\end{eqnarray}

\paragraph{i. Expansion of $\hat\psi^b(\theta_0)$ and $\hat\psi_\theta^b(\theta_0)$:}
\begin{align*}
  \hat{\psi}^b(\theta^b) &=
  \psi(\theta^b)+\frac{\mymathcal A^b(\theta^b)}{\sqrt{T}}+
    \frac{\mymathcal C^b(\theta^b)}{T}+o_p(\frac{1}{T})\\
      &= \psi(\theta_0+ \frac{A^b(\theta_0)}{\sqrt{T}}+\frac{C^b(\theta_0)}{T}+o_p(\frac{1}{T}))
       + \frac{\mymathcal A^b(\theta_0 + \frac{A^b(\theta_0)}{\sqrt
       {T}}+\frac{C^b(\theta_0)}{T}+o_p(\frac{1}{T}))}{\sqrt{T}}\\
&  + \frac{\mymathcal C^b(\theta_0 + \frac{A^b(\theta_0)}{\sqrt{T}}
  +\frac{C^b(\theta_0)}{T}+o_p(\frac{1}{T}))}{T}+o_p(\frac{1}{T})\\
        & = \psi(\theta_0)
             + \frac{\mymathcal A^b(\theta_0)}{\sqrt{T}}+\frac{\psi_\theta(\theta_0)A^b(\theta_0)}{\sqrt{T}}
         + \frac{\mymathcal C^b(\theta_0)}{T} + \frac{\mymathcal A_\theta^b(\theta_0)A^b(\theta_0)}{T} \\ &
         +\frac{1}{2}\sum_{j=1}^K \frac{\psi_{\theta,\theta_j}
         (\theta_0)A^b(\theta_0)A^b_j(\theta_0)}{T}+o_p(\frac{1}{T}).
\end{align*}
Since $\hat \psi^b(\theta^b)$ equals $\hat \psi$ for all $b$,
\begin{eqnarray}
  A^b(\theta_0)&=&\Big[ \psi_\theta(\theta_0)\Big]^{-1} \left( \mymathcal A
  (\theta_0) - \mymathcal A^b(\theta_0)\right) \label{eq:RS-A}\\
  C^b(\theta_0)&=&\Big[ \psi_\theta(\theta_0)\Big]^{-1} \left( \mymathcal C
  (\theta_0)-\mymathcal C^b(\theta_0)-\frac{1}{2}\sum_{j=1}^K \psi_{\theta,\theta_j}
  (\theta_0)A^b(\theta_0)A^b_j(\theta_0) - \mymathcal A_\theta^b(\theta_0)A^b
  (\theta_0)\right),\label{eq:RS-C}
\end{eqnarray}
it follows that
\begin{eqnarray*}
\hat\psi^b_\theta(\theta^b)
        &=&\hat\psi^b_\theta\bigg(\theta_0+\frac{A^b(\theta_0)}{\sqrt{T}}+\frac
        {C^b(\theta_0)} {T}+o_p(\frac{1}{T})\bigg)\\
        &=&\psi_\theta\bigg(\theta_0+\frac{A^b(\theta_0)}{\sqrt{T}}+\frac{C^b(\theta_0)} {T}+o_p(\frac{1}{T})\bigg)
        + \frac{\mymathcal A^b_\theta\bigg(\theta_0+\frac{A^b(\theta_0)}{\sqrt{T}}+\frac{C^b(\theta_0)} {T}+o_p(\frac{1}{T})\bigg)}{\sqrt{T}}\\
        &&+ \frac{\mymathcal C^b_\theta\bigg(\theta_0+\frac{A^b(\theta_0)}{\sqrt{T}}+\frac{C^b(\theta_0)} {T}+o_p(\frac{1}{T})\bigg)}{T} + o_p(\frac{1}{T})\\
        &=&\psi_\theta(\theta_0)+ \sum_{j=1}^K \frac{\psi_{\theta,\theta_j}(\theta_0) A_j^b(\theta_0)}{\sqrt{T}}+\frac{\mymathcal A^b_\theta(\theta_0)}{\sqrt{T}}
        + \frac{1}{2}\sum_{j=1}^K\sum_{k=1}^K\frac{\psi_{\theta,\theta_j,\theta_k}
        (\theta_0)A_j^b(\theta_0)A_k^b(\theta_0)}{T}\\ &&+\sum_{j=1}^K\frac{\psi_{\theta,\theta_j}(\theta_0)C^b_j(\theta_0)}{T}
        + \sum_{j=1}^K \frac{\mymathcal A^b_{\theta,\theta_j}(\theta_0)A^b_j(\theta_0)}{T} + \frac{\mymathcal C^b(\theta_0)}{T}+o_p(\frac{1}{T}).
\end{eqnarray*}
To obtain the determinant of $\hat\psi^b_\theta(\theta^b)$, let   $a^b(\theta_0)=\text{trace}(\mathcal{A}^b(\theta_0))$,
  $a^b_2(\theta_0)=\text{trace}(\mathcal{A}^b(\theta_0)^2)$,
  $c^b(\theta_0)=\text{trace}(\mathcal{C}^b(\theta_0))$, where
\begin{align*}
  &\mathcal{A}^b(\theta_0)=\Big[ \psi_\theta(\theta_0) \Big]^{-1}\left( \sum_{j=1}^K \psi_{\theta,\theta_j}(\theta_0) A_j^b(\theta_0)+\mymathcal A^b_\theta(\theta_0) \right)\\
  &\mathcal{C}^b(\theta_0)=\Big[ \psi_\theta(\theta_0) \Big]^{-1}\Bigg(\frac
  {1}{2}\sum_{j=1}^K\sum_{k=1}^K\frac{\psi_{\theta,\theta_j,\theta_k}(\theta_0)A_j^b(\theta_0)A_k^b(\theta_0)}{T}+\sum_{j=1}^K\frac{\psi_{\theta,\theta_j}(\theta_0)C^b_j(\theta_0)}{T}+\sum_{j=1}^K \mymathcal A^b_{\theta,\theta_j}(\theta_0)A^b_j(\theta_0) + \mymathcal C^b(\theta_0) \Bigg).
\end{align*}
Now for any matrix $X$ with all eigenvalues smaller than $1$ we have: $\log
(I_K+X)=X-\frac{1}{2}X^2+o(X)$. Furthermore, for any matrix $M$ the determinant
$|M|=\exp(\text{trace}(\log M)))$. Together, these imply that for arbitrary
$X_1,X_2$:
\begin{align*}
  \Big|I+\frac{X_1}{\sqrt{T}}+\frac{X_2}{T} + o_p(\frac{1}{T})\Big| &=  \exp \left( \text{trace}\left( \frac{X_1}{\sqrt{T}}+\frac{X_2}{T} + \frac{X_1^2}{T} + o_p(\frac{1}{T}) \right) \right) \\&= 1+  \frac{\text{trace}\left(X_1\right)}{\sqrt{T}}+\frac{\text{trace}\left(X_2\right)}{T}+ \frac{\text{trace}\left(X_1^2\right)}{T} + o_p(\frac{1}{T}).
\end{align*}
Hence the required determinant is
\begin{align*}
  \Big| \hat\psi^b_\theta(\theta^b) \Big|
  = \Big| \hat\psi_\theta(\theta_0) \Big|\Big| I + \frac{\mathcal{A}^b(\theta_0)}{\sqrt{T}} + \frac{\mathcal{C}^b(\theta_0)}{T} + o_p(\frac{1}{T})\Big|
  = \Big| \hat\psi_\theta(\theta_0) \Big| \left( 1 + \frac{a^b(\theta_0)}{\sqrt{T}} + \frac{a^b_2(\theta_0)}{T}+ \frac{c^b(\theta_0)}{T} +o_p(\frac{1}{T})\right).
\end{align*}

\paragraph{ii. Expansion of  $w^b(\theta^b)=|\hat\psi_\theta(\theta^b)|^{-1}\pi
(\theta^b)$:}
\begin{align*}
  \Big| \hat\psi^b_\theta(\theta^b) \Big|^{-1} \pi(\theta^b)
  &= \Big| \hat\psi_\theta(\theta_0) \Big|^{-1} \left( 1 + \frac{a^b(\theta_0)}{\sqrt{T}} + \frac{a^b_2(\theta_0)}{T}+ \frac{c^b(\theta_0)}{T} +o_p(\frac{1}{T})\right)^{-1}
\pi(\theta_0 + \frac{A^b(\theta_0)}{\sqrt{T}}+\frac{C^b(\theta_0)}{T}+o_p(\frac{1}{T}))\\
&= \Big| \hat\psi_\theta(\theta_0) \Big|^{-1} \left( 1 - \frac{a^b(\theta_0)}{\sqrt{T}} - \frac{a^b_2(\theta_0)}{T} - \frac{c^b(\theta_0)}{T} +o_p(\frac{1}{T})\right)\\&\times \left( \pi(\theta_0) +\pi_\theta(\theta_0)\frac{A^b(\theta_0)}{\sqrt{T}}+\pi_\theta(\theta_0)\frac{C^b(\theta_0)}{T}+\frac{1}{2}\sum_{j=1}^K\frac{\pi_{\theta,\theta_j}(\theta_0)A^b(\theta_0)A^b_j(\theta_0)}{T}+o_p(\frac{1}{T}) \right)
\\
&= \Big| \hat\psi_\theta(\theta_0) \Big|^{-1} \pi(\theta_0) \Bigg( 1-\frac{a^b(\theta_0)}{\sqrt{T}}+\frac{\pi_\theta(\theta_0)}{\pi(\theta_0)}\frac{A^b(\theta_0)}{\sqrt{T}}
-\frac{a_2^b(\theta_0)}{T}-\frac{c^b(\theta_0)}{T}\\&-\frac{\pi_\theta(\theta_0)}{\pi(\theta_0)}\frac{a^b(\theta_0)A^b(\theta_0)}{T}+\frac{\pi_\theta(\theta_0)}{\pi(\theta_0)}\frac{C^b(\theta_0)}{T}+\frac{1}{2}\frac{A^b(\theta_0)\pi_{\theta,\theta^\prime}(\theta_0)A^{b \prime}(\theta_0)}{T}+o_p(\frac{1}{T}) \Bigg).
\end{align*}
Now $\bar A(\theta_0) = \frac{1}{B}\sum_{b=1}^B A^b(\theta_0)$.
Similarly define $\bar C(\theta_0)=\frac{1}{B} C^b(\theta_0)$.
Also, denote the term in $1/T$ by: \[e^b(\theta_0)=-a_2^b(\theta_0)-c^b(\theta_0)-\frac{\pi_\theta(\theta_0)}{\pi(\theta_0)}a^b(\theta_0)A^b(\theta_0)+\frac{\pi_\theta(\theta_0)}{\pi(\theta_0)}C^b(\theta_0)+\frac{1}{2}A^b(\theta_0)\pi_{\theta,\theta^\prime}(\theta_0)A^{b \prime}(\theta_0).\]

The normalized weight for draw $b$ is:
\begin{align*}
\bar w^b(\theta^b)&= \frac{ \Big| \hat\psi^b_\theta(\theta^b) \Big|^{-1} \pi(\theta^b)}{\sum_{c=1}^B \Big| \hat\psi^c_\theta(\theta^c) \Big|^{-1} \pi(\theta^c)} =
  \frac{1}{B}\bigg(\frac{1-\frac{a^b(\theta_0)}{\sqrt{T}}+\frac{\pi_\theta
  (\theta_0)}{\pi(\theta_0)}\frac{A^b(\theta_0)}{\sqrt{T}}+\frac{e^b(\theta_0)}
  {T}+o_p(\frac{1}{T})}{1+\frac{1}{B}\sum_{c=1}^B -\frac{a^c(\theta_0)}{\sqrt
  {T}}+\frac{\pi_\theta(\theta_0)}{\pi(\theta_0)}\frac{A^c(\theta_0)}{\sqrt{T}}+\frac{e^c(\theta_0)}{T}+o_p(\frac{1}{T})} \bigg)\\
  &= \frac{1}{B}\bigg(\frac{1-\frac{a^b(\theta_0)}{\sqrt{T}}+\frac{\pi_\theta
  (\theta_0)}{\pi(\theta_0)}\frac{A^b(\theta_0)}{\sqrt{T}}+\frac{e^b(\theta_0)}
  {T}+o_p(\frac{1}{T})}{1-\frac{\bar a(\theta_0)}{\sqrt{T}}+\frac{\pi_\theta
  (\theta_0)}{\pi(\theta_0)}\frac{\bar A(\theta_0)}{\sqrt{T}}+\frac{\bar e(\theta_0)}{T}+o_p(\frac{1}{T})}\bigg) \\
  &= \frac{1}{B}\Big(1-\frac{a^b(\theta_0)}{\sqrt{T}}+\frac{\pi_\theta(\theta_0)}{\pi(\theta_0)}\frac{A^b(\theta_0)}{\sqrt{T}}+\frac{e^b(\theta_0)}{T}+o_p(\frac{1}{T})\Big) \times\Big( 1+\frac{\bar a(\theta_0)}{\sqrt{T}}-\frac{\pi_\theta(\theta_0)}{\pi(\theta_0)}\frac{\bar A(\theta_0)}{\sqrt{T}}-\frac{\bar e(\theta_0)}{T}+o_p(\frac{1}{T})\Big) \\
  &= \frac{1}{B}\Big( 1-\frac{a^b(\theta_0)-\bar a(\theta_0)}{\sqrt{T}}+\frac{\pi_\theta(\theta_0)}{\pi(\theta_0)}\frac{A^b(\theta_0)-\bar A(\theta_0)}{\sqrt{T}}
  + \frac{e^b(\theta_0)-\bar e(\theta_0)}{T}-\frac{a^b(\theta_0)\bar a(\theta_0)}
  {T}-\frac{\pi_\theta(\theta_0)}{\pi(\theta_0)}\frac{A^b(\theta_0)\bar a
  (\theta_0)}{T}\\ & \;-\frac{\pi_\theta(\theta_0)}{\pi(\theta_0)}\frac{\bar A(\theta_0) a^b(\theta_0)}{T} - \left[ \frac{\pi_\theta(\theta_0)\pi_\theta(\theta_0)^\prime}{\pi(\theta_0)^2} \right] \frac{A^b(\theta_0)^\prime\bar A(\theta_0)}{T}+o_p(\frac{1}{T})\Big).
\end{align*}
The posterior mean is $\bar \theta_{RS}=\sum_{b=1}^B\bar w^b(\theta^b)\theta^b$. Using
 $\theta^b$ defined in (\ref{eq:A2}), $A$ and $C$ defined in (\ref{eq:RS-A}) and (\ref{eq:RS-C}):
 \begin{align*}
\bar \theta_{RS} &=  \theta_0 + \frac{1}{B}\sum_{b=1}^B\frac{ A^b(\theta_0)}{\sqrt{T}}
+ \frac{1}{B}\sum_{b=1}^B\frac{ C^b(\theta_0)}{T} + \frac{\pi_\theta(\theta_0)}{\pi(\theta_0)}
\frac {1}{B}\sum_{b=1}^B \frac{(A^b(\theta_0)-\bar A(\theta_0))A^b(\theta_0)}
{T}+C^M(\theta_0)+o_p(\frac{1}{T}).
\end{align*}

\setcounter{equation}{0}
\setcounter{section}{0}
\renewcommand{\thesection}{B.\arabic{section}}
\renewcommand{\theequation}{B.\arabic{equation}}

\section{Proof of Results for LT }
From
\begin{align*}
  &\theta^b = \theta_0 + \frac{A^b(\theta_0)}{\sqrt{T}}+\frac{C^b(\theta_0)}{T}+o_p(\frac{1}{T}),
\end{align*}
we have, given that $\hat \psi_b$ is drawn from the asymptotic distribution of $\hat \psi$
\begin{align*}
  \hat{\psi}^b(\theta^b) &= \psi(\theta^b)+\frac{\mymathcal A^b_\infty(\theta^b)}{\sqrt{T}}\\
                         &= \psi\bigg(\theta_0+ \frac{A^b(\theta_0)}{\sqrt{T}}+\frac
                         {C^b(\theta_0)}{T}+o_p(\frac{1}{T})\bigg)+
                          \frac{\mymathcal A^b_\infty(\theta_0 + \frac{A^b(\theta_0)}{\sqrt{T}}+\frac{C^b(\theta_0)}{T}+o_p(\frac{1}{T}))}{\sqrt{T}}\\
                         &= \psi(\theta_0)
                         + \frac{\mymathcal A^b_\infty(\theta_0)}{\sqrt{T}}+\frac{\psi_\theta(\theta_0)A^b(\theta_0)}{\sqrt{T}}
                           \frac{\mymathcal A_{\infty,\theta}^b(\theta_0)A^b(\theta_0)}{T} +\frac{1}{2}\sum_{j=1}^K \frac{\psi_{\theta,\theta_j}(\theta_0)A^b(\theta_0)A^b_j(\theta_0)}{T}+o_p(\frac{1}{T})
\end{align*}
which is equal to $\hat \psi$ for all $b$. Hence
\begin{eqnarray}
  A^b(\theta_0)&=&\Big[ \psi_\theta(\theta_0)\Big]^{-1} \left( \mymathcal A
  (\theta_0) - \mymathcal A_\infty^b(\theta_0)\right) \label{eq:LT-A}\\
  C^b(\theta_0)&=&\Big[ \psi_\theta(\theta_0)\Big]^{-1} \left( \mymathcal C
  (\theta_0)-\frac{1}{2}\sum_{j=1}^K \psi_{\theta,\theta_j}(\theta_0)A^b(\theta_0)A^b_j(\theta_0)
  - \mymathcal A_{\infty,\theta}^b(\theta_0)A^b(\theta_0)\right). \label{eq:LT-C}
\end{eqnarray}
Note that the bias term $C^b$ depends on the bias term $\mymathcal C$. For
the weights, we need to consider
\begin{eqnarray*}
  \hat \psi^b_\theta(\theta^b) &=& \psi_\theta\left( \theta_0 + \frac{A^b(\theta_0)}{\sqrt{T}} + \frac{C^b(\theta_0)}{T} + o_p(\frac{1}{T}) \right) + \frac{\mymathcal{A}^b_{\infty,\theta}\left( \theta_0 +\frac{A^b(\theta_0)}{\sqrt{T}} + \frac{C^b(\theta_0)}{T}+o_p(\frac{1}{T}) \right)}{\sqrt{T}} \\
  &=& \psi_\theta(\theta_0) + \sum_{j=1}^K \frac{\psi_{\theta,\theta_j}(\theta_0)A^b_j(\theta_0)}
  {\sqrt{T}} + \frac{\mymathcal{A}^b_{\infty,\theta}(\theta_0)}{\sqrt{T}} + \sum_{j=1}^k
  \frac{\psi_{\theta,\theta_j}(\theta_0)C^b_j(\theta_0)}{T} + \sum_{j=1}^K \frac{\mymathcal{A}^b_{\infty,\theta,\theta_j}A_j^b(\theta_0)}{T} \\ &&+ \frac{1}{2}\sum_{j,k=1}^K \frac{\psi_{\theta,\theta_j,\theta_k}(\theta_0)A_j^b(\theta_0)A_k^b(\theta_0)}{T}+o_p(\frac{1}{T}).
\end{eqnarray*}
Let
\begin{align*}
  &\mathcal{A}^b(\theta_0) = \Big[ \psi_\theta(\theta_0)\Big]^{-1}\left( \mymathcal{A}^b_{\infty,\theta}(\theta_0) + \sum_{j=1}^K \psi_{\theta,\theta_j}(\theta_0)A^b_j(\theta_0) \right)\\
  &\mathcal{C}^b(\theta_0) = \Big[ \psi_\theta(\theta_0) \Big]^{-1} \left( \sum_{j=1}^K \psi_{\theta,\theta_j}(\theta_0)C^b_j(\theta_0) + \sum_{j=1}^K \mymathcal{A}^b_{\infty,\theta,\theta_j}(\theta_0)A_j^b(\theta_0) + \frac{1}{2}\sum_{j=1}^K \sum_{k=1}^K \psi_{\theta,\theta_j,\theta_k}(\theta_0)A^b_j(\theta_0)A^b_k(\theta_0) \right)\\
  &a^b(\theta_0) = \text{trace}(\mathcal{A}^b(\theta_0)),  \quad a^b_2(\theta_0)
  = \text{trace}(\mathcal{A}^b(\theta_0)^2), \quad
  c^b(\theta_0) = \text{trace}(\mathcal{C}^b(\theta_0)).
\end{align*}
The determinant is
\begin{align*}
  \Big| \hat \psi^b_\theta(\theta_0)\Big|^{-1} &= \Big| \psi_\theta(\theta_0)\Big|^{-1} \Big| I + \frac{\mathcal{A}^b(\theta_0)}{\sqrt{T}} + \frac{\mathcal{C}^b(\theta_0)}{T}+o_p(\frac{1}{T})\Big|^{-1}
  = \Big| \psi_\theta(\theta_0)\Big|^{-1}\left( 1+\frac{a^b(\theta_0)}{\sqrt{T}} + \frac
  {a^b_2(\theta_0)}{T} + \frac{c^b(\theta_0)}{T} + o_p(\frac{1}{T})\right)^{-1}\\
 & = \Big| \psi_\theta(\theta_0)\Big|^{-1} \left( 1 - \frac{a^b(\theta_0)}{\sqrt{T}} - \frac{a^b_2(\theta_0)}{T} - \frac{c^b(\theta_0)}{T} + o_p(\frac{1}{T})\right).
\end{align*}
The prior is
\begin{align*}
  \pi(\theta^b) &= \pi \left( \theta_0 + \frac{A^b(\theta_0)}{\sqrt{T}} + \frac{C^b(\theta_0)}{T} + o_p(\frac{1}{T})\right) \\
  &= \pi(\theta_0) + \pi_\theta(\theta_0) \frac{A^b(\theta_0)}{\sqrt{T}} + \pi_\theta(\theta_0) \frac{C^b(\theta_0)}{T} + \frac{1}{2} \frac{A^b(\theta_0) \pi_{\theta,\theta^\prime}A^{b \prime}(\theta_0)}{T} + o_p(\frac{1}{T}).
\end{align*}
Let:
$ e^b(\theta_0) = -c^b(\theta_0) - a_2^b(\theta_0) + \frac{\pi_\theta(\theta_0)}{\pi(\theta_0)}C^b(\theta_0) + A^b(\theta_0)
\frac{\pi_{\theta,\theta^\prime}}{\pi}(\theta_0)A^{b \prime}(\theta_0). $
After some simplification, the product is
\begin{align*}
  \Big| \hat \psi^b_\theta(\theta_0)\Big|^{-1} \pi(\theta^b)
  &= \Big| \psi_\theta(\theta_0)\Big|^{-1}  \pi(\theta_0)\Big( 1-\frac{a^b(\theta_0)}{\sqrt{T}} + \frac{\pi_\theta(\theta_0)}{\pi(\theta_0)}\frac{A^b(\theta_0)}{\sqrt{T}} + \frac{e^b(\theta_0)}{T} + o_p(\frac{1}{T})\Big).
\end{align*}
Hence, the normalized weight for draw $b$ is
\begin{align*}
\bar  w^b(\theta^b) &= \frac{\Big| \hat \psi^b_\theta(\theta_0)\Big|^{-1} \pi(\theta^b)}{\sum_{c=1}^B \Big| \hat \psi^c_\theta(\theta_0)\Big|^{-1} \pi(\theta^c)}
  = \frac{1}{B} \frac{1-\frac{a^b(\theta_0)}{\sqrt{T}}+\frac{\pi_\theta(\theta_0)}{\pi(\theta_0)}\frac{A^b(\theta_0)}{\sqrt{T}}  + \frac{e^b(\theta_0)}{T} + o_p(\frac{1}{T})}{1-\frac{\bar a(\theta_0)}{\sqrt{T}}+\frac{\pi_\theta(\theta_0)}{\pi(\theta_0)}\frac{\bar A(\theta_0)}{\sqrt{T}}  + \frac{\bar e(\theta_0)}{T} + o_p(\frac{1}{T})} \\
  &= \frac{1}{B} \left(1-\frac{a^b(\theta_0)}{\sqrt{T}}+\frac{\pi_\theta(\theta_0)}
  {\pi(\theta_0)}\frac{A^b(\theta_0)}{\sqrt{T}}  + \frac{e^b(\theta_0)}{T} + o_p(\frac{1}{T})\right) \left(1+\frac{\bar a(\theta_0)}{\sqrt{T}}-\frac{\pi_\theta(\theta_0)}{\pi(\theta_0)}\frac{\bar A(\theta_0)}{\sqrt{T}}  - \frac{\bar e(\theta_0)}{T} + o_p(\frac{1}{T})\right)\\
   & = \frac{1}{B} \Big( 1 - \frac{a^b(\theta_0) - \bar a(\theta_0)}{\sqrt{T}} + \frac{\pi_\theta(\theta_0)}{\pi(\theta_0)} \frac{A^b(\theta_0) - \bar A(\theta_0)}{\sqrt{T}} + \frac{e^b(\theta_0)-\bar e(\theta_0)}{T} -  \frac{a^b(\theta_0) \bar a(\theta_0) }{T} - \frac{\frac{\pi_\theta(\theta_0)}{\pi(\theta_0)}A^b(\theta_0)\frac{\pi_\theta(\theta_0)}{\pi(\theta_0)}\bar A(\theta_0)}{T} \\&+\frac{\pi_\theta(\theta_0)}{\pi(\theta_0)}\frac{a^b(\theta_0)\bar A(\theta_0)}{T}+\frac{\pi_\theta(\theta_0)}{\pi(\theta_0)}\frac{\bar a(\theta_0) A^b(\theta_0)}{T} + o_p(\frac{1}{T})\Big).
\end{align*}
Hence the posterior mean is   $\bar \theta_\text{LT} = \sum_{b=1}^B \bar
w^b(\theta^b)\theta^b $ and $\theta^b=\left( \theta_0 + \frac{A^b(\theta_0)}{\sqrt{T}}
+ \frac{C^b(\theta_0)}{T} + o_p(\frac{1}{T})\right)$. After simplification, we have

\begin{align*}
  \bar \theta_\text{LT}
  &= \theta_0 + \frac{\bar A(\theta_0)}{\sqrt{T}} + \frac{\bar C(\theta_0)}{T} - \frac
  {1}{B}\sum_{b=1}^B \frac{(a^b(\theta_0)-\bar a(\theta_0))A^b(\theta_0)}{T} - \frac{[\frac{\pi_\theta(\theta_0)}{\pi(\theta_0)}\bar A(\theta_0)]^2 \theta_0}{T} + \frac{1}{B}\frac{\pi_\theta(\theta_0)}{\pi(\theta_0)} \sum_{b=1}^B \frac{(A^b(\theta_0)-\bar A(\theta_0))A^b(\theta_0)}{T} \\&- \frac{\bar{a}(\theta_0)^2\theta_0}{T}
   + 2 \frac{\pi_\theta(\theta_0)}{\pi(\theta_0)} \frac{\bar a(\theta_0) \bar A(\theta_0) \theta_0}
   {T}+ o_p(\frac{1}{T})\\
   &=  \theta_0 + \frac{ \bar A(\theta_0)}{\sqrt{T}}
+ \frac{\bar C(\theta_0)}{T} + \frac{\pi_\theta(\theta_0)}{\pi(\theta_0)}
\frac {1}{B}\sum_{b=1}^B \frac{(A^b(\theta_0)-\bar A(\theta_0))A^b(\theta_0)}
{T}+C^M(\theta_0)+o_p(\frac{1}{T}),
\end{align*}
where all terms are based on $A^b(\theta_0)$ defined in (\ref{eq:LT-A}) and
$C^b (\theta_0)$ in (\ref{eq:LT-C}).


\setcounter{equation}{0}
\setcounter{section}{0}
\renewcommand{\thesection}{C.\arabic{section}}
\renewcommand{\theequation}{C.\arabic{equation}}

\section{Results for SLT:}
From
\begin{align*}
  &\hat\psi^b(\theta) = \frac{1}{S}\sum_{s=1}^S \hat \psi^s(\theta) + \frac{\mymathcal{A}_\infty^b(\theta)}{\sqrt{T}} \\
  &\hat \psi^s(\theta)= \psi(\theta)+\frac{\mymathcal{A}^s(\theta)}{\sqrt{T}}+\frac{\mymathcal{C}^s(\theta)}{T}+o_p(\frac{1}{T})\\
  &\theta^b = \theta_0 +\frac{A^b(\theta_0)}{\sqrt{T}}+\frac{C^b(\theta_0)}{T}+o_p(\frac{1}{T}),
\end{align*}
we have
\begin{align*}
  \hat \psi^s(\theta^b) &=  \frac{1}{S}\sum_{s=1}^S \hat \psi^s\bigg(\theta_0 +\frac{A^b(\theta_0)}{\sqrt{T}}+\frac{C^b(\theta_0)}{T}+o_p(\frac{1}{T})\bigg) + \frac{\mymathcal{A}_\infty^b(\theta_0 +\frac{A^b(\theta_0)}{\sqrt{T}}+\frac{C^b(\theta_0)}{T}+o_p(\frac{1}{T}))}{\sqrt{T}} \\
  &= \psi(\theta_0)+\frac{1}{S}\sum_{s=1}^S \frac{\mymathcal{A}^s(\theta_0)}
  {\sqrt{T}} + \frac{\mymathcal{A}_\infty^b(\theta_0)}{\sqrt{T}} + \psi_\theta(\theta_0)\frac
  {A^b(\theta_0)}{\sqrt{T}} + \frac{1}{S}\sum_{s=1}^S \frac{\mymathcal{A}^s_{\theta}(\theta_0)A^b(\theta_0)}{T}+\frac{\mymathcal{A}^b_{\infty,\theta}(\theta_0)A^b(\theta_0)}{T}\\ &+\frac{1}{S}\sum_{s=1}^S \frac{\mymathcal{C}^s(\theta_0)}{T}
  +\frac{1}{2}\sum_{j=1}^K \psi_{\theta,\theta_j}(\theta_0) \frac{A^b(\theta_0)A^b_j(\theta_0)}
  {T}+\psi_\theta(\theta_0)\frac{C^b(\theta_0)}{T}+o_p(\frac{1}{T}).
\end{align*}
Thus, {\small
\begin{eqnarray}
  A^b(\theta_0) &=& \Big[ \psi_\theta(\theta_0)\Big]^{-1} \left( \mymathcal
  {A}(\theta_0) - \frac{1}{S}\sum_{s=1}^S \mymathcal{A}^s(\theta_0)-\mymathcal
  {A}_\infty^b(\theta_0)\right) \label{eq:SLT-A}\\
  C^b(\theta_0) &=& \Big[ \psi_\theta(\theta_0) \Big]^{-1}\left( \mymathcal
  {C}(\theta_0)-\frac{1}{S}\sum_{s=1}^S \mymathcal{C}^s(\theta_0) - \frac
  {1}{2}\sum_{j=1}^K \psi_{\theta,\theta_j}(\theta_0)A^b(\theta_0)A^b_j(\theta_0)\right)\nonumber
\\ &&  - \Big[ \psi_\theta(\theta_0) \Big]^{-1}\left[\frac
  {1}{S}\sum_{s=1}^S \mymathcal{A}_\theta^s(\theta_0)+\mymathcal{A}_{\infty,\theta}^b
  (\theta_0) \right]A^b(\theta_0).  \label{eq:SLT-C}
\end{eqnarray}
}
Note that we have $\mymathcal{A}_\infty^b \sim \mathcal{N}$ while $\mymathcal{A}^s \overset{d}{\to} \mathcal{N}$.
To compute the weight for draw $b$, consider
\begin{align*}
  \hat \psi^b(\theta^b) &= \psi_\theta\bigg( \theta_0 +\frac{A^b(\theta_0)}{\sqrt
  {T}}+\frac{C^b(\theta_0)}{T}+o_p(\frac{1}{T}) \bigg) + \frac{1}{S}\sum_{s=1}^S\frac{\mymathcal{A}^s\bigg(\theta_0 +\frac{A^b(\theta_0)}{\sqrt{T}}+\frac{C^b(\theta_0)}{T}+o_p(\frac{1}{T})\bigg)}{\sqrt{T}}\\
  &+ \frac{\mymathcal{A}_\infty^b\bigg(\theta_0 +\frac{A^b(\theta_0)}{\sqrt{T}}+\frac{C^b(\theta_0)}{T}+o_p(\frac{1}{T})\bigg)}{\sqrt{T}} + \frac{1}{S}\sum_{s=1}^S \frac{\mymathcal{C}^s\bigg(\theta_0 +\frac{A^b(\theta_0)}{\sqrt{T}}+\frac{C^b(\theta_0)}{T}+o_p(\frac{1}{T})\bigg)}{T}+o_p(\frac{1}{T}) \\
  &= \psi_\theta(\theta_0) + \sum_{j=1}^K \psi_{\theta,\theta_j}(\theta_0)\frac{A_j^b(\theta_0)}{\sqrt{T}}+\frac{1}{S}\sum_{s=1}^S\frac{\mymathcal{A}^s_\theta(\theta_0)}{\sqrt{T}}+\frac{\mymathcal{A}^b_{\infty,\theta}(\theta_0)}{\sqrt{T}} + \frac{1}{S}\sum_{s=1}^S \frac{\mymathcal{C}^s(\theta_0)}{T} + \sum_{j=1}^K
  \psi_{\theta,\theta_j}(\theta_0)\frac{C^b_j(\theta_0)}{T} \\&+ \frac{1}{S} \sum_{s=1}^S \sum_{j=1}^K \frac{\mymathcal{A}_{\theta,\theta_j}^s(\theta_0)A^b_j(\theta_0)}{T}+\sum_{j=1}^K\frac{\mymathcal{A}_{\infty,\theta,\theta_j}^b(\theta_0)A^b_j(\theta_0)}{T} +  \frac{1}{2}\sum_{j=1}^K \sum_{k=1}^K\psi_{\theta,\theta_j,\theta_k}(\theta_0)\frac{A^b_k(\theta_0)A^b_j(\theta_0)}{T}+o_p(\frac{1}{T}).
\end{align*}
Let:
\begin{align*}
  \mathcal{A}^b(\theta_0) &= \Big[ \psi_\theta(\theta_0) \Big]^{-1} \left( \frac{1}{S}\sum_{s=1}^S\mymathcal{A}^s_\theta(\theta_0)+\mymathcal{A}^b_{\infty,\theta}(\theta_0) + \sum_{j=1}^K\psi_{\theta,\theta_j}A^b_j(\theta_0) \right)\\
  \mathcal{C}^b(\theta_0) &= \Big[ \psi_\theta(\theta_0) \Big]^{-1} \left(
  \frac{1}{S}\sum_{s=1}^S \mymathcal{C}^s(\theta_0) + \sum_{j=1}^K \Big[\psi_{\theta,\theta_j}
  (\theta_0)C^b_j(\theta_0) + \frac{1}{S}\sum_{s=1}^S\mymathcal{A}^s_{\theta,\theta_j}
  (\theta_0)A^b_j(\theta_0) + \mymathcal{A}^b_{\infty,\theta,\theta_j}(\theta_0)A^b_j
  (\theta_0)\Big]\right)\\ &+\Big[ \psi_\theta(\theta_0) \Big]^{-1}
  \left(\frac{1}{2}\sum_{j,k=1}^K \psi_{\theta,\theta_j,\theta_k}(\theta_0)A_k^b(\theta_0)A_j^b(\theta_0)\right)\\
  &a^b(\theta_0)  =\text{trace}(\mathcal{A}^b(\theta_0)), \quad
  a^b_2(\theta_0)=\text{trace}(\mathcal{A}^b(\theta_0)^2),\quad
  c^b(\theta_0)  =\text{trace}(\mathcal{C}^b(\theta_0)).
\end{align*}
The determinant is
\begin{align*}
  \Big| \hat \psi^b(\theta^b)\Big|^{-1} = \Big| \psi_\theta(\theta_0)\Big|^{-1} \left( 1- \frac{a^b(\theta_0)}{\sqrt{T}}-\frac{a_2^b(\theta_0)}{T}-\frac{c^b(\theta_0)}{T}+o_p(\frac{1}{T})\right).
\end{align*}
Hence
\begin{align*}
  \Big| \hat \psi^b(\theta^b)\Big|^{-1}\pi(\theta^b) &=
  \Big| \psi_\theta(\theta_0) \Big|^{-1}\pi(\theta_0)\left( 1-\frac{a^b(\theta_0)}{\sqrt{T}}-\frac{a_2^b(\theta_0)}{T}-\frac{c^b(\theta_0)}{T}+o_p(\frac{1}{T})\right)\\ &\times\left( 1+\frac{\pi_\theta(\theta_0)}{\pi(\theta_0)}\frac{A^b(\theta_0)}{\sqrt{T}} + \frac{\pi_\theta(\theta_0)}{\pi(\theta_0)}\frac{C^b(\theta_0)}{T}+\frac{1}{2}\sum_{j=1}^K \frac{\pi_{\theta,\theta_j}(\theta_0)}{\pi(\theta_0)}\frac{A^b(\theta_0)A^b_j(\theta_0)}{T}+o_p(\frac{1}{T})\right)\\
  &= \Big| \psi_\theta(\theta_0) \Big|^{-1}\pi(\theta_0) \left( 1-\frac{a^b(\theta_0)}{\sqrt{T}}+\frac{\pi_\theta(\theta_0)}{\pi(\theta_0)}\frac{A^b(\theta_0)}{\sqrt{T}}+ \frac{e^b(\theta_0)}{T}+o_p(\frac{1}{T})\right)
\end{align*}
where $e^b(\theta_0) = -a^b(\theta_0) \frac{\pi_\theta(\theta_0)}{\pi(\theta_0)}A^b(\theta_0)-a^b_2(\theta_0) - c^b(\theta_0) +\frac{\pi_\theta(\theta_0)}{\pi(\theta_0)}C^b(\theta_0)+\frac{1}{2}\sum_{j=1}^K \frac{\pi_{\theta,\theta_j}(\theta_0)}{\pi(\theta_0)}A^b(\theta_0)A^b_j(\theta_0).$
The normalized weights are
\begin{align*}
\bar  w^b(\theta^b) &= \frac{  \Big| \hat \psi^b(\theta^b)\Big|^{-1}\pi(\theta^b)}{\sum_{c=1}^B   \Big| \hat \psi^c(\theta^c)\Big|^{-1}\pi(\theta^c)}
  \\&= \frac{1}{B}\left( 1-\frac{a^b(\theta_0)}{\sqrt{T}}+\frac{\pi_\theta(\theta_0)}{\pi(\theta_0)}\frac{A^b(\theta_0)}{\sqrt{T}}+\frac{e^b(\theta_0)}{T}+o_p(\frac{1}{T}) \right)
  \left( 1+\frac{\bar a(\theta_0)}{\sqrt{T}}-\frac{\pi_\theta(\theta_0)}{\pi(\theta_0)}\frac{\bar A(\theta_0)}{\sqrt{T}}-\frac{\bar e(\theta_0)}{T}+o_p(\frac{1}{T}) \right). \\
\end{align*}
The posterior mean   $\bar\theta_{\text{SLT}} = \sum_{b=1}^B \bar w^b(\theta^b)\theta^b
$ with $\theta^b=\theta_0 +\frac{A^b(\theta_0)}{\sqrt{T}}+\frac{C^b(\theta_0)}{T}+o_p(\frac{1}
{T})$. After some simplification,

\begin{eqnarray*}
\bar \theta_{SLT}  &=&
  \theta_0 + \frac{\bar A(\theta_0)}{\sqrt{T}} + \frac{\bar C(\theta_0)}{T} + \frac{\pi_\theta(\theta_0)}{\pi(\theta_0)}\frac{1}{B}\sum_{B=1}^B\frac{(A^b(\theta_0)-\bar A(\theta_0))A^b(\theta_0)}{T} - \frac{1}{B}\sum_{b=1}^B\frac{(a^b(\theta_0)-\bar a(\theta_0))A^b(\theta_0)}{T}
  \\&&+2 \frac{\pi_\theta(\theta_0)}{\pi(\theta_0)}\frac{\bar a(\theta_0) \bar A(\theta_0) \theta_0}
  {T}-\frac{\bar a^2(\theta_0) \theta_0}{T}-[\frac{\pi_\theta(\theta_0)}{\pi(\theta_0)}\bar A(\theta_0)]^2\frac{\theta_0}{T}+o_p(\frac{1}{T})\\
  &=&
  \theta_0 + \frac{\bar A(\theta_0)}{\sqrt{T}} + \frac{\bar C(\theta_0)}{T} + \frac{\pi_\theta
  (\theta_0)}{\pi(\theta_0)}\frac{1}{B}\sum_{B=1}^B\frac{(A^b(\theta_0)-\bar A(\theta_0))A^b(\theta_0)}
  {T} +C^M(\theta_0)+o_p(\frac{1}{T})
\end{eqnarray*}
where terms in $A$ and $C$  are defined from (\ref{eq:SLT-A}) and (\ref{eq:SLT-C}).

\newpage
\setcounter{equation}{0}
\setcounter{section}{0}
\renewcommand{\thesection}{D.\arabic{section}}
\renewcommand{\theequation}{D.\arabic{equation}}

\section{Results For The Example in Section 6.1}
The data generating process is $y_t= m_0+\sigma_0 e_t$, $e_t\sim iid\;\mathcal N
(0,1)$.  As a matter of notation, a hat is used to denote the mode, a
bar denotes the mean, superscript
$s$ denotes a specific draw and  a  subscript $S$ to denote average over $S$ draws.
For example,   $\bar e_S=\frac
{1}{ST} \sum_{s=1}^S \sum_{t=1}^T e_t^s=\frac{1}{S}\sum_{s=1}^S \bar
e^s$.
\paragraph{MLE:} Define $\bar e=\frac{1}{T}\sum_{t=1}^T e_t$. Then the mean estimator
is $\hat
m= m_0+
\sigma_0\bar
e\sim N(0,\sigma_0^2/T)$.
For the variance estimator,  $\hat e= y-\hat m =\sigma_0(e - \bar e)=
\sigma_0 Me$, $M=I_T-1(1^\prime 1)^{-1} 1^\prime$ is an idempotent
matrix with $T-1$ degrees of freedom. Hence
$\hat \sigma^2_
{ML} = \hat e^\prime \hat e /T \sim
\sigma_0^2 \chi^2 _ {T-1}$.

\paragraph{BC:} Expressed in terms of sufficient statistics $(\hat m,\hat\sigma^2)$,
the joint density of $\mathbf y$ is
\[ p(\mathbf y; m,\sigma^2)= (\frac{1}{2\pi\sigma^2})^{T/2} \exp \bigg (-\frac{\sum_
{t=1}^T (m-\hat m)^2}{2\sigma^2 } \times  \frac{-T\hat\sigma^2}{2 \sigma^2} \bigg).
\]
The flat prior is $\pi(m,\sigma^2)\propto1$. The marginal posterior
distribution for $\sigma^2$ is $p(\sigma^2|\mathbf y) =\int _{-\infty}^\infty p(\mathbf
y|m,\sigma^2) dm$. Using the result that $\int_{-\infty}^\infty \exp(-\frac
{T}{2\sigma^2} (m-\hat m)^2) dm=\sqrt{2\pi\sigma^2}$, we have
\begin{eqnarray*}
 p(\sigma^2|\mathbf y) &\propto&  (2\pi\sigma^2)^{-(T-1)/2} \exp(-T
\hat\sigma^2/2\sigma^2) \sim \text{inv}\Gamma\bigg(\frac{T-3}{2},\frac{T\hat\sigma^2}{2}\bigg).
\end{eqnarray*}
The mean of an $\text{inv}\Gamma(\alpha,\beta)$ is $\frac{\beta}{\alpha-1}$. Hence
the BC posterior is $\bar \sigma^2_{BC}=E(\sigma^2|\mathbf y) = \hat\sigma^2 \frac
{T} {T-5}$.
\paragraph{SMD:}   The estimator
equates the  auxiliary statistics computed from the sample
with the average of the statistics over simulations. Given $\sigma$, the
mean estimator $\hat m_S$ solves
$\hat m=
\hat m_S+ \sigma \frac{1}{S}\sum_{s=1}^S  \bar e^s $. Since we use sufficient statistics,
$\hat m$ is the ML estimator. Thus, $\hat m_S\sim \mathcal N(m, \frac{\sigma_0^2}
{T} +\frac{\sigma^2}
{ST})$. Since $y_t^s-\bar y_t^s= \sigma(e_t^s-\bar e^s)$,
 the variance estimator $\hat \sigma^2_S$ is the $\sigma^2$ that solves $\hat
\sigma^2 =\sigma^2 (\frac {1} {ST} \sum_ {s=1}^S
\sum_{t=1}^T (e_t^s-\bar e^s)^2)$
Hence
\[\hat \sigma^2_S = \frac{\hat\sigma^2}{\frac{1}{ST}\sum_s\sum_t (\hat
e_t^s-\bar
e^s)^2}=\sigma^2\frac{\chi^2_{T-1}/T}{\chi^2_{S(T-1)}/(ST)}=\sigma^2
F_{T-1,S(T-1)}.\]
The mean of a $F_{d_1,d_2}$ random variable is  $\frac{d_2}{d_2-2}$. Hence $E(\hat\sigma^2_
{SMD})= \sigma^2\frac{(T-1)}{S(T-1)-2}$.

\paragraph{LT:}
The LT is defined as
\[ p_\text{LT} (\sigma^2 |\hat \sigma^2) \propto \mathbbm{1}_{\sigma^2 \geq 0}\exp\left(-\frac{T}{2}\frac{\left( \hat \sigma^2 - \sigma^2\right)^2}{2\hat \sigma^4}\right) \]
which implies
\[ \sigma^2 |\hat \sigma^2 \sim_{\text{LT}} \mathcal{N}\left( \hat \sigma^2, \frac{2\hat \sigma^4}{T} \right) \text{ truncated to } [0,+\infty[.\]
For $X\sim \mathcal{N}(\mu, \sigma^2)$ we have $\mathbb{E}(X|X>a)=\mu + \frac{\phi(\frac{a-\mu}{\sigma})}{1-\Phi(\frac{a-\mu}{\sigma})}\sigma$ (Mills-Ratio). Hence:
\begin{align*}
  \mathbb{E}_\text{LT}(\sigma^2|\hat \sigma^2) &= \hat \sigma^2 + \frac{\phi(\frac{0-\hat \sigma^2}{\sqrt{2/T}\hat \sigma^2})}{1-\Phi(\frac{0-\hat \sigma^2}{\sqrt{2/T}\hat \sigma^2})}\sqrt{2/T}\hat \sigma^2
  = \hat \sigma^2 \left(1+\sqrt{\frac{2}{T}}\frac{\phi(-\sqrt{T/2})}{1-\Phi(-\sqrt{T/2})} \right).
\end{align*}
Let $\kappa_\text{LT}=\sqrt{\frac{2}{T}}\frac{\phi(-\sqrt{T/2})}{1-\Phi(-\sqrt{T/2})}$. We have
$
  \mathbb{E}_\text{LT}(\sigma^2|\hat \sigma^2) = \hat \sigma^2 \left(1+\kappa_\text{LT} \right).
$
The expectation of the estimator is
\[ \mathbb{E}\left( \mathbb{E}_\text{LT}(\sigma^2|\hat \sigma^2) \right) = \sigma^2 \frac{T-1}{T} \left(1+\kappa_\text{LT}  \right) \]
from which we deduce the bias of the estimator
\[ \mathbb{E}\left( \mathbb{E}_\text{LT}(\sigma^2|\hat \sigma^2) \right) - \sigma^2 = \sigma^2  \left(\frac{T-1}{T}\kappa_\text{LT} -\frac{1}{T}\right). \]
The variance of the estimator is $ 2\sigma^4 \frac{T-1}{T^2} \left(1+\kappa_\text{LT} \right)^2$
and the Mean-Squared Error (MSE)
\[ \sigma^4 \left( 2\frac{T-1}{T^2} \left(1+\kappa_\text{LT} \right)^2 + \left(\frac{T-1}{T}\kappa_\text{LT}  -\frac{1}{T}\right)^2 \right) \]
which is the squared bias of MLE plus terms that involve the Mills-Ratio (due to the truncation).

\paragraph{SLT:}
The SLT is defined as
\[ p_\text{SLT} (\sigma^2 |\hat \sigma^2) \propto \mathbbm{1}_{\sigma^2 \geq 0}\exp\left(-\frac{T}{2}\frac{\left( \hat \sigma^2 - \sigma^2\frac{\chi^2_{S(T-1)}}{ST} \right)^2}{2\hat \sigma^4}\right) = \mathbbm{1}_{\sigma^2 \geq 0}\exp\left(-\frac{T[\frac{\chi^2_{S(T-1)}}{ST}]^2}{2}\frac{\left( \hat \sigma^2/\frac{\chi^2_{S(T-1)}}{ST} - \sigma^2 \right)^2}{2\hat \sigma^4}\right)\]
where
\[ \hat \sigma^2_S = \sigma^2 \frac{1}{S}\sum_{s=1}^2\frac{1}{T}\sum_{t=1}^T(e_t^s-\bar e^s)^2 = \sigma^2 \frac{\chi^2_{S(T-1)}}{ST}. \]
This yields the slightly more complicated formula
\[ \sigma^2 |\hat \sigma^2, (e^s)_{s=1,\dots,S}  \sim \mathcal{N}\left( \hat \sigma^2/\frac{\chi^2_{S(T-1)}}{ST}, \frac{2 \hat \sigma^4}{T}[\frac{ST}{\chi^2_{S(T-1)}}]^2 \right)\]
and the posterior mean becomes
\begin{align*}
  \mathbb{E}_\text{SLT}(\sigma^2|\hat \sigma^2) &= \hat \sigma^2\frac{ST}{\chi^2_{S(T-1)}} +\frac{\phi\left(-\frac{\hat \sigma^2 ST/\chi^2_{S(T-1)}}{\sqrt{\frac{2\hat \sigma^4}{T}(\frac{ST}{\chi^2_{S(T-1)}})^2}}\right)}{1-\Phi\left(-\frac{\hat \sigma^2 ST/\chi^2_{S(T-1)}}{\sqrt{\frac{2\hat \sigma^4}{T}(\frac{ST}{\chi^2_{S(T-1)}})^2}}\right)}\sqrt{2/T}\frac{ST}{\chi^2_{S(T-1)}}\hat \sigma^2 \\
  &= \hat \sigma^2\frac{ST}{\chi^2_{S(T-1)}} +\frac{\phi\left(-\sqrt{T/2}\right)}{1-\Phi\left(-\sqrt{T/2}\right)}\sqrt{2/T}\frac{ST}{\chi^2_{S(T-1)}}\hat \sigma^2.
\end{align*}
Let $\kappa_\text{SLT}=\frac{\phi(-\sqrt{T/2})}{1-\Phi(-\sqrt{T/2})}\sqrt{2/T}\frac{ST}{\chi^2_{S(T-1)}}=\kappa_\text{LT} \frac{ST}{\chi^2_{S(T-1)}}$ (random).
We can compute
\[ \mathbb{E} \left( \mathbb{E}_\text{SLT}(\sigma^2|\hat \sigma^2) \right) = \sigma^2 \frac{S(T-1)}{S(T-1)-2} + \sigma^2 \frac{T-1}{T}\mathbb{E}(\kappa_\text{SLT})\]
and the bias
\[ \mathbb{E} \left( \mathbb{E}_\text{SLT}(\sigma^2|\hat \sigma^2) \right) - \sigma^2 = \sigma^2 \frac{2}{S(T-1)-2} + \sigma^2 \frac{T-1}{T} \mathbb{E}(\kappa_\text{SLT})\]
which is the bias of SMD and the Mills-Ratio term that comes from taking the mean of the truncated normal rather than the mode.
The variance is similar to the LT and the SMD
\begin{align*}
  2\sigma^4\kappa_1\frac{1}{T-1} + 2\sigma^4 \mathbb{V}(\kappa_\text{SLT}) +4\sigma^4 \frac{T-1}{T^2}\text{Cov}(\kappa_\text{SLT},\frac{S}{\chi^2_{S(T-1)}}).
\end{align*}
The extra term is due to $\kappa_\text{SLT}$ being random. We could simplify further noting that
   $\kappa_\text{SLT}=\kappa_\text{LT}\frac{ST}{\chi^2_{S(T-1)}}$,
   $\mathbb{E}(\kappa_\text{SLT})=\kappa_{\text{LT}}\frac{ST}{S(T-1)-2}$,
   $\mathbb{V}(\kappa_\text{SLT})=\kappa_\text{LT}^2\frac{S^2T^2}{(S(T-1)-2)^2(S(T-1)-4)}$ and
   $\text{Cov}(\kappa_\text{SLT},\frac{S}{\chi^2_{S(T-1)}}) = \kappa_\text{LT}S^2T\mathbb{V}(1/\chi^2_{S(T-1)})=\kappa_\text{LT}\frac{S^2T}{(S(T-1)-2)^2(S(T-1)-4)}.$\\

The MSE is
\begin{align*}
  \sigma^4\left[  \frac{2}{S(T-1)-2} + \frac{T-1}{T}\mathbb{E}(\kappa_\text{SLT}) \right]^2 +2\sigma^4\kappa_1\frac{1}{T-1} + 2\sigma^4 \mathbb{V}(\kappa_\text{SLT}) +4\sigma^4 \frac{T-1}{T^2}\text{Cov}(\kappa_\text{SLT},\frac{S}{\chi^2_{S(T-1)}}) \\ =
  \underbrace{2\sigma^4\left[  \frac{2}{[S(T-1)-2]^2} +\kappa_1\frac{1}{T-1} \right]}_{\text{MSE of SMD}} +  \frac{(T-1)^2}{T^2}\mathbb{E}(\kappa_\text{SLT}^2+ \frac{4\sigma^4}{S(T-1)-2}\frac{T-1}{T}\mathbb{E}(\kappa_\text{SLT}) \\+  2\sigma^4 \mathbb{V}(\kappa_\text{SLT}) +4\sigma^4 \frac{T-1}{T^2}\text{Cov}(\kappa_\text{SLT},\frac{S}{\chi^2_{S(T-1)}})  .
\end{align*}

\paragraph{RS:} The auxiliary statistic for each draw of simulated data
is matched to the sample auxiliary statistic. Thus,
$\hat m=m^b+\sigma^b \bar e^b$. Thus conditional on $\hat m$
and $\sigma^{2,b}$, $m^b=\hat m-\sigma^b \bar e^b \sim \mathcal
N(0,\sigma^{2,b}/T)$. For the variance, $\hat{\sigma}^{2,b}=\sigma^{2,b} \sum_t
(e_t^b-\bar e^b)^2/T$. Hence
\[ \sigma^{2,b} = \frac{\hat \sigma^2}{\sum_t (e_t^b-\bar e^b)^2/T} =
\sigma^2 \frac{\sum_t (e_t -\bar e)^2/T}{ \sum_t (e_t^b-\bar
e^b)^2/T}\sim \text{inv}\Gamma\bigg(\frac{T-1}{2}, \frac{T\hat\sigma^2}
{2}\bigg)\]
Note that $p_{BC}(\sigma^2|\hat{\sigma}^2) \sim \text{inv}\Gamma\bigg(\frac{T-3}{2}, \frac{T\hat\sigma^2}
{2}\bigg)$ under a flat prior, the Jacobian adjusts to the posterior to match the true posterior.
To compute the posterior mean, we need to compute the Jacobian of the transformation:
$|\psi_\theta|^{-1}=\frac{\partial \sigma^{2,s}}{\partial
\hat\sigma^2}$\footnote{This holds because $\hat \sigma^{2,b}(\sigma^{2,b})=\hat\sigma^2$ so that $|d\hat \sigma^{2,b}/d\sigma^{2,b}|^{-1}=|d \sigma^{2,b}/d\hat \sigma^{2}|$.}. Since $\sigma^{2,b}=\frac{ T\hat\sigma^2}{\sum_t (e_t^b-\bar e^b)^2}$,
$|\psi_\theta|^{-1}= \frac {T} {\sum_t (e_t^b-\bar e^b)^2}$.

Under the prior $p(\sigma^{2,s}) \propto 1$, the posterior mean without the Jacobian transformation
is
\[ \bar \sigma^2 = \sigma^2 \frac{1}{B}\sum_{b=1}^B
\frac{\sum_t (e_t -\bar e)^2/T}{ \sum_t (e_t^b-\bar
e^b)^2/T} \overset{B \to \infty}{\longrightarrow} \hat{\sigma}^2\frac{T}{T-3}
\]
The  posterior mean after adjusting for the Jacobian transformation is
\begin{eqnarray*}
\bar\sigma^2_{RS}&=&
\frac{\sum_{b=1}^B \sigma^{2,b}\cdot \frac{T}{\sum_t
(e_t^b-\bar e^b)^2}} {\sum_
{b=1}^B 1/\sigma^ {2,b}}
= \hat\sigma^2\frac {\sum_b (\frac{T}{\sum_t (e_t^b-\bar e^b)^2} )^2 }
{\sum_{b=1}\sum_t (e_t^b-\bar e^b)^2/T}
= T\hat \sigma^2 \frac{\frac{1}{B}\sum_b (z^b)^2}{\frac{1}{B}\sum_b
z^b}\end{eqnarray*}
where $1/z^b=\sum_t (e_t^b-\bar e^b)^2$.
As $B\rightarrow\infty$, $\frac{1}{B}\sum_b (z^b)^2\pconv
 E[(z^{b})^2]$ and $\frac{1}{S}\sum_b z^b\pconv E[z^b]$.
Now  $z^b\sim \text
{inv}\chi^2_{T-1}$ with mean $\frac{1}{T-3}$ and variance $\frac{2}{
(T-3)^2(T-5)}$ giving $E[(z^{b})^2] = \frac{1}{(T-3)(T-5)}$. Hence as
$B\rightarrow\infty$, $\bar
\sigma^2_{RS ,R}= \hat\sigma^2  \frac{T}{T-5}=\bar \sigma^2_
{BC}$.

\paragraph{Derivation of the Bias Reducing Prior}
The bias of the MLE estimator has $ \mymathcal{E}(\hat{\sigma}) = \sigma^2 - \frac{1}{T}\sigma^2
$  and variance $V(\hat{\sigma}^2) = 2\sigma^4(\frac{1}{T}-\frac{1}{T^2})$.
Since the auxiliary parameters coincide with the parameters of interest, $\nabla_
\theta \psi(\theta)$ and $\nabla_{\theta\theta^\prime}\psi(\theta)=0$.
For $Z \sim \mathcal{N}(0,1), A(v;\sigma^2)=\sqrt{2}\sigma^2(1-\frac{1}{T})Z$,
Thus $\partial_{\sigma^2} A(v;\sigma^2) = \sqrt{2}(1-\frac{1}{T})Z, a^s =
\sqrt{2}\sigma^2(1-\frac{1}{T})(Z-Z^s)$. The terms in the asymptotic expansion are
therefore
\begin{eqnarray*}
\partial_{\sigma^2} A(v^s;\sigma^2) a^s &=& 2\sigma^2(1-\frac{1}{T})^2Z^s(Z-Z^s)
\Rightarrow \mymathcal{E}(\partial_{\sigma^2} A(v^s;\sigma^2) a^s) = -\sigma^2 2(1-\frac{1}{T})^2
\\
V(a^s) &=& 4\sigma^4(1-\frac{1}{T})^2\\
 cov(a^s,a^{s'}) &=& 2(1-\frac{1}{T})^2\sigma^4
\\
(1-\frac{1}{S})V(a^s) + \frac{S-1}{S} cov(a^s,a^{s'}) &=&
\sigma^4(1-\frac{1}{T})^2 \Big( 4(1-\frac{1}{S}) + 2\frac{S-1}{S} \Big) = \frac{\sigma^
2 S }{3(S-1)}
\end{eqnarray*}
Noting that $|\partial_{\hat\sigma^2}\sigma^{2,b}| \propto \sigma^{2,b}$,
it is analytically simpler in this example to solve for the weights directly,
i.e. $w(\sigma^2)=\pi(\sigma^2)|\partial_{\hat\sigma^2}\sigma^{2,b}|$ rather
than the bias reducing prior $\pi$ itself. Thus the bias reducing prior satisfies
\[ \partial_{\sigma^2} w(\sigma^2) =
\frac{-2\sigma^2(1-\frac{1}{T})^2}{\sigma^4(1-\frac{1}{T})^2 \Big(
4(1-\frac{1}{S}) + 2\frac{S-1}{S} \Big)} =
-\frac{1}{\sigma^2}\frac{2}{4(1-\frac{1}{S}) + 2\frac{S-1}{S}}.\]
Taking the integral on both sides we get:
\[\log(w(\sigma^2)) \propto -\log(\sigma^2)  \Rightarrow w(\sigma^2)
\propto \frac{1}{\sigma^2}  \Rightarrow \pi(\sigma^2)
\propto \frac{1}{\sigma^4} \]
which is the Jeffreys prior if there is no re-weighting and the square of the Jeffreys prior when we use the Jacobian to re-weight. Since the estimator for the mean was unbiased,
 $\pi(m) \propto 1$ is the prior for $m$.

The  posterior mean under the Bias Reducing Prior $\pi(\sigma^{2,s})=1/\sigma^{4,s}$ is the same as the posterior without weights but using the Jeffreys prior  $\pi(\sigma^{2,s})=1/\sigma^{2,s}$:
\begin{eqnarray*}
\bar\sigma^2_{RS }&=&
\frac{\sum_{s=1}^S \sigma^{2,s} (1/\sigma^{2,s})}{\sum_{s=1}^S 1/\sigma^{2,s}}
=\frac {S}{\sum_{s=1}^S 1/\sigma^{2,s}}
= \sigma^2 \frac{\sum_{t=1}^T (e_t-\bar e)^2/T}{\sum_{s=1}^S
\sum_{t=1}^T (e_t^s-\bar e^s)^2/(ST)}\equiv \hat\sigma^2_{SMD}.
\end{eqnarray*}

\newpage
\section{Further Results for Dynamic Panel Model with Fixed Effects}
\begin{table}[ht]
  \caption{Dynamic Panel $\rho=0.9,\beta=1,\sigma^2=2$}
\label{tbl:table2.2}

\begin{center}
Mean over 1000 replications

\begin{tabular}{rrrrrrrrrrr}
 \hline \hline
& & MLE  & LT & SLT & SMD & ABC  & RS & Bootstrap \\
 \hline \hline
 & Mean & 0.751 & 0.751 & 0.895 & 0.898 & 0.889 & 0.899 & 0.751 \\
   $\hat \rho:$ & SD & 0.030 & 0.030 & 0.026 & 0.025 & 0.025 & 0.025 & 0.059 \\
   & Bias & -0.149 & -0.149 & -0.005 & -0.002 & -0.011 & -0.001 & -0.149 \\ \hline \hline
   & Mean & 0.934 & 0.934 & 0.998 & 1.000 & 0.996 & 1.000 & 0.935 \\
   $\hat \beta:$&SD & 0.070 & 0.071 & 0.074 & 0.073 & 0.073 & 0.073 & 0.139 \\
   &Bias & -0.066 & -0.066 & -0.002 & 0.000 & -0.004 & 0.000 & -0.065 \\  \hline \hline
   &Mean &  1.857 & 1.865 & 1.972 & 1.989 & 2.054 & 2.097 & 1.858 \\
   $\hat \sigma^2:$ &SD &  0.135 & 0.141 & 0.145 & 0.145 & 0.151 & 0.153 & 0.269 \\
   &Bias &  -0.143 & -0.135 & -0.028 & -0.011 & 0.054 & 0.097 & -0.142 \\  \hline \hline
    S & & -- & -- & 500 & 500 & 1 & 1 & 500\\
    B & & -- & 500 & 500 & -- & 500 & 500 & -- \\
  \hline \hline
\end{tabular}\\ \end{center}
See note to Table 3.

\end{table}
\newpage

\clearpage
\baselineskip=12.0pt
\normalsize
\bibliography{metrics,kn,macro,metrics2}

\end{document}